\documentclass[aps,prl,floatfix,twocolumn,superscriptaddress]{revtex4}
\usepackage{latexsym}
\usepackage{graphicx}
\usepackage{rotating}
\usepackage{amsmath,amssymb,amsfonts}
\usepackage[colorlinks=true,linkcolor=blue,citecolor=red]{hyperref}
\usepackage{color}

\newcommand{\dz}{d$_{3z^{2}-r^{2}}$ }
\newcommand{\dx}{d$_{x^{2}-y^{2}}$ }

\newcommand{\dxz}{d$_{xz}$ }
\newcommand{\dzd}{d$_{3z^{2}-r^{2}}$}

\newcommand{\dxyd}{d$_{xy}$}
\newcommand{\dyzd}{d$_{yz}$}
\newcommand{\dxzd}{d$_{xz}$}

\newcommand{\emphasize}{\emph}

\def\onlinecite#1{\cite{#1}}

\newcommand{\cred}{\color{black}}

\newcommand{\cblack}{\color{black}}

\usepackage{natbib}
\usepackage[compact]{titlesec}
\titlespacing{\section}{0pt}{*0}{*0}
\titlespacing{\subsection}{0pt}{*0}{*0}
\titlespacing{\subsubsection}{0pt}{*0}{*0}
\begin{document}

\title{Importance of many body effects in the kernel of 
\cred hemoglobin \cblack
for ligand binding}

\author{C\'edric Weber}
\affiliation{King's College London, Theory and Simulation of Condensed Matter (TSCM), The Strand, London WC2R 2LS}
\affiliation{Cavendish Laboratory,  J. J. Thomson Av., Cambridge CB3 0HE, U.K.}
\author{David D. O'Regan}
\affiliation{Cavendish Laboratory,  J. J. Thomson Av., Cambridge CB3 0HE, U.K.}
\affiliation{Theory and Simulation of Materials, \'{E}cole Polytechnique F\'{e}d\'{e}rale de Lausanne, Station 12, 1015 Lausanne, Switzerland}
\author{Nicholas D. M. Hine}
\affiliation{Cavendish Laboratory,  J. J. Thomson Av., Cambridge CB3 0HE, U.K.}
\affiliation{Department of Materials, Imperial College London, Exhibition Road, London SW7 2AZ, U.K.}
\author{Peter B. Littlewood}
\affiliation{Cavendish Laboratory,  J. J. Thomson Av., Cambridge CB3 0HE, U.K.}
\affiliation{Physical Sciences and Engineering, Argonne National Laboratory, Argonne, Illinois 60439, U.S.A.}
\author{Gabriel Kotliar}
\affiliation{Rutgers University, 136 Frelinghuysen Road, Piscataway, NJ, U.S.A.}
\author{Mike C. Payne}
\affiliation{Cavendish Laboratory,  J. J. Thomson Av., Cambridge CB3 0HE, U.K.}


\begin{abstract}
We propose a mechanism for 
binding of diatomic ligands to heme based
on a dynamical orbital selection process. 
This scenario may be described as 
\emphasize{bonding determined by local 
valence fluctuations}.
We support this model using 
linear-scaling first-principles calculations, 
in combination with dynamical mean-field theory, 
applied to heme, the kernel of the
\cred hemoglobin \cblack 
metalloprotein central to human respiration.
\cred
We find that variations in 
Hund's exchange coupling 
induce a reduction of the iron $3d$ density, with a concomitant increase of valence fluctuations. \cblack
We discuss the comparison between our computed
 optical absorption spectra and experimental data,
our picture accounting for the observation of
optical transitions in the infrared regime,
and how the Hund's coupling reduces, by a factor of five, the
strong imbalance in the binding energies of heme with 
CO and O$_2$ ligands.
\end{abstract}

\maketitle

Metalloporphyrin systems, such as heme, 
play a central role in biochemistry. 
The ability of such molecules
to reversibly bind small ligands is of great interest, 
particularly in the case of heme which binds diatomic
ligands such as oxygen and carbon monoxide. 
Heme acts as a transport
molecule for oxygen in human respiration, 
while carbon monoxide inhibits this function.
Despite intensive studies \cite{PhysRevB.79.245404,PhysRevB.82.081102,heme_dft_u_surface}, the
binding of the iron atom at centre of the heme molecule to O$_2$ and CO ligands
remains poorly understood.
In particular, one problem obtained with density functional theory \cite{kohn_sham} (DFT) approaches
to ligand binding of heme 
is that the difference in the binding energy ($\Delta \Delta E$) of carboxy-heme and oxy-heme 
is very large, and the theory predicts an unrealistic binding affinity to CO, several orders of magnitude larger than to O$_2$ \cite{heme_to_bind_or_not,heme_angle_binding_co}.

  Recent progress has been made to cure this problem 
 using  DFT+$U$ \cred for molecular systems~\cite{heme_marzari,
PhysRevLett.97.103001}, with which it was found that the
inclusion of many body effects in the calculations reduced the imbalance between O$_2$ and CO affinities~\cite{cole}. 
\cblack
Inclusion of conformal modifications, such as the Fe-C-O binding angle  \cite{heme_angle_fe_o}, or 
the deviation of the Fe atom from the porphyrin plane, were also shown to affect CO and O$_2$ binding energies.

A general problem encountered by DFT is the 
strong dependence of the 
energetics and the spin state on small changes in the geometry. 
In particular, traditional DFT fails to describe the correct high-spin 
ground state of heme molecules. 
\cred DFT+$U$ provides an improved description \cite{heme_marzari, dftu_heme_FeCl}, \cblack
but is known to overestimate magnetic moments and gives often artificial and non physical
spin-symmetry-broken states. Moreover, the 
\cred rotationally-invariant \cblack
DFT+$U$ methodology does not capture 
well the effect of the Hund's coupling $J$,
which is known to be large in iron based systems. 
It was recently shown that the effect of 
strong correlations are not always 
driven by the Coulomb repulsion $U$ alone, but in some cases act in combination with the Hund's coupling $J$ \cite{spin_disorder_ref_gabi_zipping,luca_hund_coupling,luca_prl_hund_janus}. 
Understanding the effect of strong correlations in heme, and in particular how the symmetry of the highest occupied molecular orbital (HOMO) is affected by $U$ and $J$, is important in
the context of describing the CO binding, which was shown to be strongly dependent on the HOMO symmetry ~\cite{add_ref_co_homo_lumo}.

Recent progress has been made in this direction by dynamical mean field theory ~\cite{OLD_GABIS_REVIEW} (DMFT), combined with DFT (DFT+DMFT)
which can refine the description of the charge and spin of correlated ions, and describes in a remarkable way the strong correlations, 
induced by both $U$ and $J$. Also, DFT can only describe a static magnetic moment associated with a spin symmetry
broken state, and requires the inclusion of the spin-orbit interaction to explain a change of spin states \cite{heme_paper_oxy_deoxy_heme}.
This is not necessary at the DMFT level, which describes both static and fluctuating magnetic moments within the same framework.

In this work, we extend the DFT+$U$ analysis by means
of the combination of state-of-the-art 
linear scaling DFT~\cite{onetep_ref_ngwf} with DMFT,
and apply this methodology to heme. 
The methodology builds upon our earlier works ~\cite{our_paper_vo2}
and is described in detail in the supplementary material.

Although DFT+DMFT has been widely used to study solids, 
in this study we apply
our real-space 
DFT+DMFT implementation 
to a moderately large molecule, extending the 
scope of applicability of DMFT to biology in an unprecedented manner. 
DMFT allows the quantum and thermal fluctuations, missing in
zero-temperature DFT calculations, to be recovered.
Moreover, it includes within the calculation both the Coulomb repulsion $U$
and the Hund's coupling $J$. Which of $U$ or $J$ drives the
 many body effects in heme \cite{luca_prl_hund_janus} remains an open question, paramount
 to understanding ligand binding, that we address in this work.   
Methods are available to obtain $U$ and $J$ 
\cred parameters appropriate to DMFT \cblack
\cite{dynamical_screening_u_silke_metal_oxides},
but in this work we focus on the dependence of the results with the Hund's coupling $J$,
and we verify that our calculations are not sensitive to the Coulomb repulsion $U$ or
to the temperature $T$ \cite{supplementary_material}. 
The key question that we address in this work is: to what extent does the Hund's coupling, so far neglected in all studies
applied to heme, affect the binding of heme to O$_2$ and CO ligands, and in particular does $J$ reduce the strong affinity for CO binding? 
If not specified otherwise, we use a similar value $U=4$~eV 
\cred 
to those previously computed for
 DFT+$U$~\cite{heme_marzari}, \cblack and ambient temperature $T=294$K.
The methodology is described in detail the supplementary material.
Ionic geometries were obtained for four different configurations:
unligated deoxyheme, FeP-d; the heme-CO complex carboxyheme, FeP(CO);
the heme-O$_2$ complex oxyheme, FeP(O$_2$); and a theoretical
planar version of deoxyheme, FeP-p.


\begin{figure}
\begin{center}
\includegraphics[width=1.0\columnwidth]{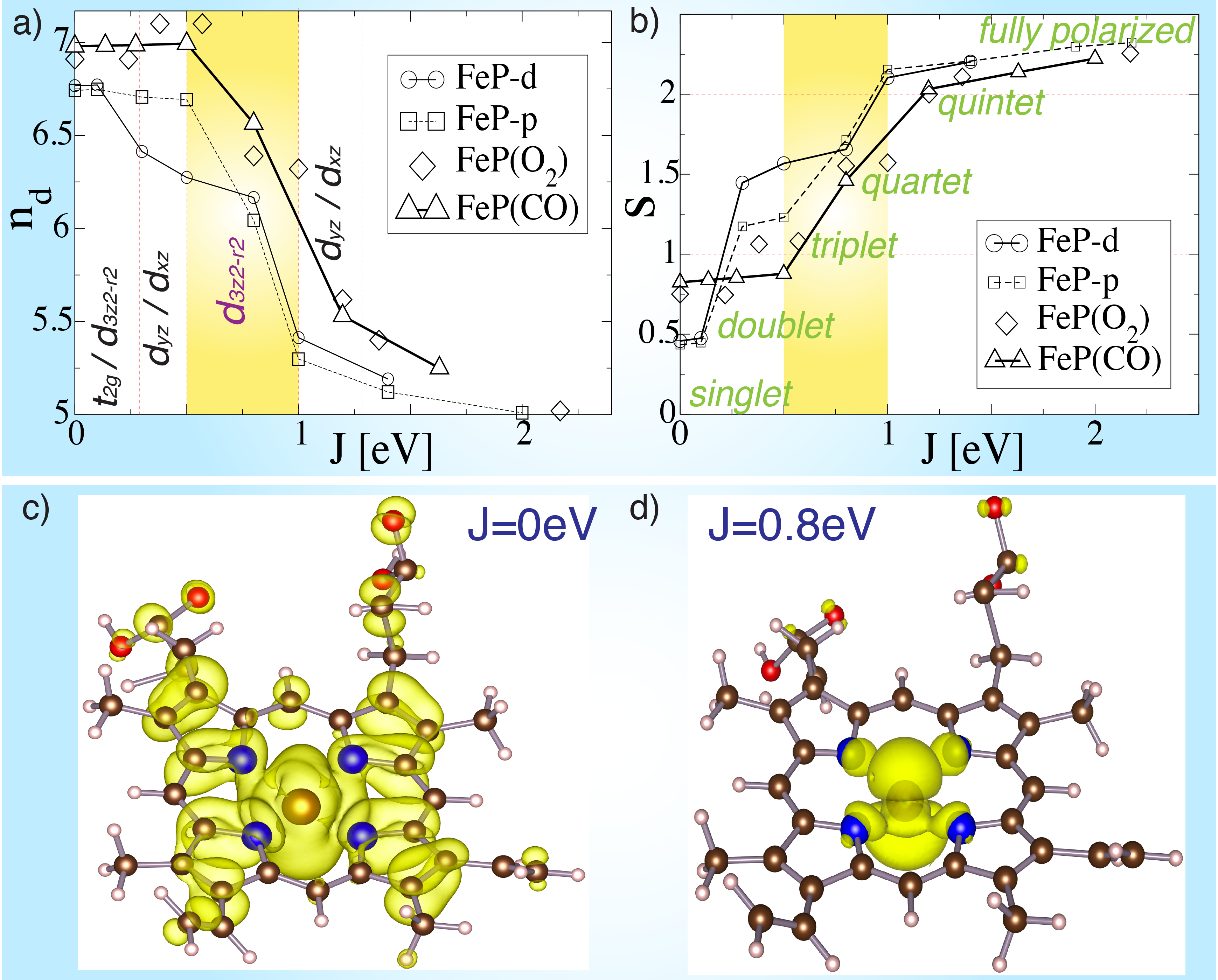}
\caption{ 
{\bf Orbital selection scenario: }  
Dependence of a) the iron $3d$ subspace occupancy $n_d$ and b) the effective spin 
quantum number $s$ on the Hund's coupling $J$, for both
unligated and ligated heme models.
The physically relevant region \cred $0.5~$eV~$<J<1$~eV 
\cblack is highlighted in yellow.
Isosurfaces of the real-space representation of the electronic 
spectral density 
of the HOMO of FeP-d for c) $J=0$~eV and d)  $J=0.8$~eV. The 
large central sphere shows the location of the iron atom, 
and the four blue spheres indicate nitrogen atoms.}
\label{fig1}
\end{center}
\end{figure}
We first discuss the dependence of the iron 
$3d$ subspace occupancy $n_d$ 
on the Hund's coupling parameter $J$ (Fig.~\ref{fig1}.\textbf{a}). 
We emphasize that the expectation value of the 
occupancy $n_d$ of the iron $3d$ sub-shell 
is not constrained to integer values in DFT and DFT+DMFT, 
since the iron occupation is a local observable, and hence does not
commute with the Hamiltonian and is not conserved and there are
valence fluctuations.
\cblack

In the typical region of 
physically meaningful values of the 
Hund's coupling for iron $3d$ electrons, 
$J \approx 0.8$~eV,~\cite{hund_value_iron_arsenide}
we find a very sharp dependence of the electronic density on $J$. 
In fact, $J \approx 0.8$~eV places heme 
directly in the transition region between low-spin states and the $n_d=5$~e 
fully-polarized state obtained for large Hund's coupling.
We note that our results are weakly dependent on the choice of the Coulomb repulsion U
(see sup. material).

In Fig.~\ref{fig1}.\textbf{b}, we show the effective quantum
spin number, which is associated to the norm of the angular spin 
vector $\bf{S}$ by the usual relation $\left| \bf{S} \right| = \sqrt{s(s+1)}$.
The spin $s$ shows characteristic plateaux as a 
function of the Hund's coupling at the semi-classically allowed values of the
magnetization (corresponding to pure doublet, triplet, quartet, 
and quintet states). A fully-polarized state is recovered 
for sufficiently large Hund's coupling, as  expected. 

At $J=0.8$~eV, and almost irrespective of ligation and doming, 
we find that heme has a spin expectation 
value of $s\approx1.5$ corresponding to a quartet  state in a semi-classical picture. 
Our results indicate that the true many-body wave-function of FeP-d is
thus an entangled superposition of triplet and quintet states.
The proposition that  heme might be in an entangled state was pointed out early \cite{porphyrin_mixed_state} in the
context of a Pariser-Parr-Pople model Hamiltonian, and is confirmed by our DMFT calculations. 
In particular, this accounts for the striking differences obtained experimentally for very similar porphyrin
systems, e.g. it was found that unligated FeP is a triplet \cite{fetpp_triplet_state} in the tetraphenylporphine configuration, a
triplet with different orbital symmetry in the octaethylporphine configuration \cite{feoep_triplet_state}, and a quintet in the 
octamethyltetrabenzporphine configuration \cite{feotbp_quintet_state}. The strong dependence of the spin state with
respect to small modifications in the structure is consistent with an entangled spin state. 

In our calculations, we find that both oxyheme and 
carboxyheme adopt a low spin state for $J<0.25$eV 
and larger multiplicities in the physical region of
$J\approx 0.8$eV, while in both cases the spin state is
very close in character to that of unligated deoxyheme. 
Significantly, we observe only subtle
differences between FeP(O$_2$), FeP(CO) 
and FeP-d for $J=0.8$~eV, 
while the DFT and DFT+$U$ treatment yields ground-states 
for carboxyheme and oxyheme of pure
closed-shell and open-shell singlet configurations, 
respectively~\cite{cole,heme_marzari,heme_angle_binding_co}. 

Moreover, we find that the symmetry of the highest occupied molecular 
orbital (HOMO) of FeP-d, 
as estimated from the real-space spectral density
of the prominent feature below the Fermi level, 
is highly dependent on the Hund's coupling $J$. 
In particular, for $J=0$~eV,  the HOMO is an admixture of 
orbital characters (see vertical labels in Fig.~\ref{fig1}.\textbf{a}). 
However, the Hund's coupling drives a rather complex orbital selection, 
such that for the region of greatest interest, $J\approx0.8$~eV, 
the HOMO predominantly exhibits \dz symmetry. 
The orbital selection process also induces a pinning 
of the Fermi density to the quantum impurity, 
such that it is delocalized for $J=0$~eV 
(see Fig.~\ref{fig1}.\textbf{c}), 
while for $J=0.8$~eV (Fig.~\ref{fig1}.\textbf{d}) 
it is instead localized to the iron $3d$ sub-shell.

In our view, this relates to the Fe-O-O angle obtained in FeP(O$_2$)  \cite{co_o2_imbalance_p_bonding_dz}. 
Indeed, the bent geometry of FeP(O$_2$) can be explained by a favorable interaction 
between the p*-orbital of the O$_2$ and the \dz -orbital on Fe \cite{co_o2_imbalance_p_bonding_dz}: 
the O$_2$ p*-orbital is closer in energy to \dz compared to the p*-orbitals in CO, and 
hence it gains more energy by bending, which increases the overlap. 
For FeP(CO) the situation is opposite, and there is no stabilization gained by bending \cite{co_o2_imbalance_p_bonding_dz}. 
On the contrary, the bending in FeP(CO) is induced by the strain of the 
protein and it reduces the binding energy.  Naively, the orbital selection of the \dz orbital
is hence expected to go in the direction of curing the strong O$_2$ and CO imbalance.
Moreover, the charge localization at the Fermi level suggests that other artificial binding between
the non-metallic atomic orbitals of heme and strong electronegative O$_2$ will not be obtained, 
and hence will protect heme from undesired charge transfer.


\begin{figure}
\begin{center}
\includegraphics[width=0.8\columnwidth]{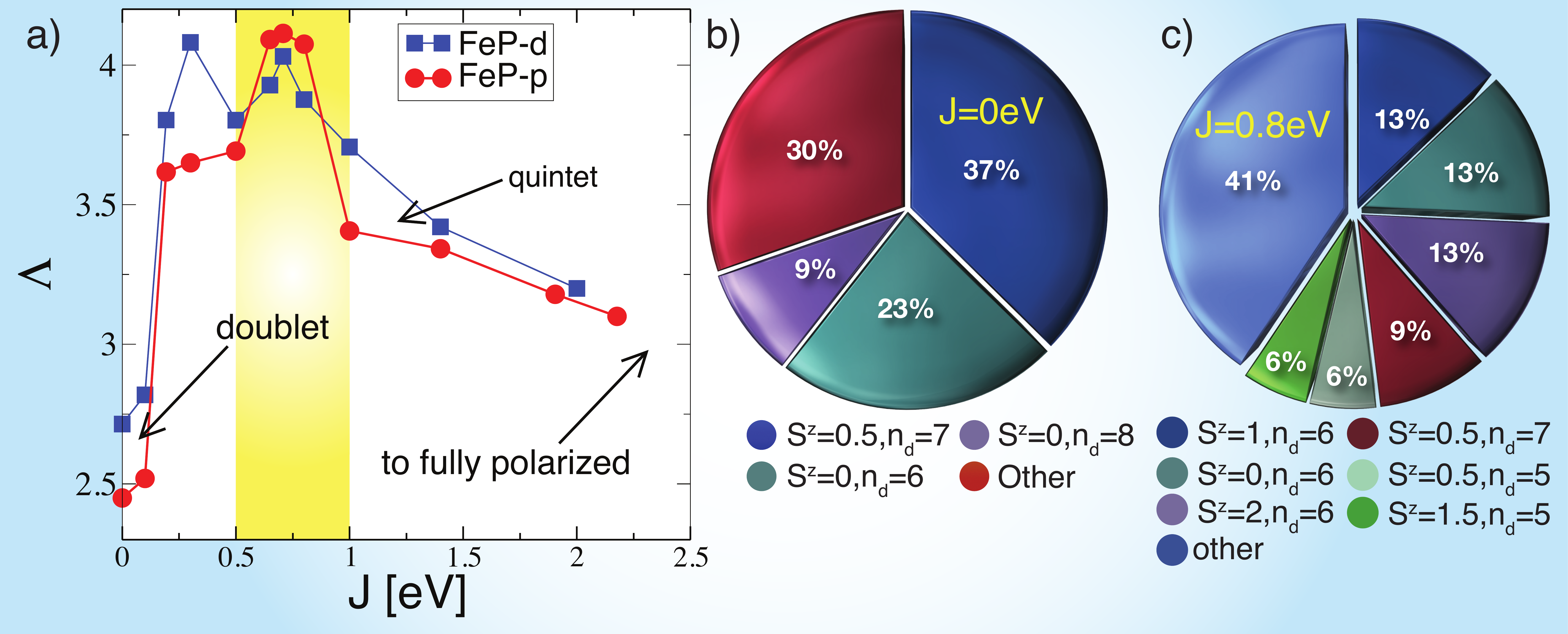}
\caption{
{\bf Valence fluctuations: } a) 
Von Neumann entropy $\Lambda$ obtained  by DFT+DMFT for FeP-p (circles) and FeP-d (squares). 
Histograms of the dominant electronic 
configurations for FeP-d
for b) $J=0$~eV and (c) $J=0.8$~eV. The pie wedge 
labelled \emphasize{other} contains configurations with a weight smaller
than $3\%$. The iron $3d$ spin $S^z$ and iron $3d$ occupancy 
$n_d$ of the dominant configurations is indicated.
}
\label{fig2}
\end{center}
\end{figure}
We now discuss the degree of quantum 
entanglement exhibited by FeP-d and FeP-p 
(see Fig.~\ref{fig2}.\textbf{a}). 
We computed the von Neumann
entropy $\Lambda=-{\rm tr} \left( \hat{\rho}_d 
\log( \hat{\rho}_d ) \right)$, 
where $\hat{\rho}_d$ 
is the reduced finite-temperature density-matrix of the iron $3d$
impurity subspace, traced over the states of the AIM bath environment. 
The entropy quantifies to what extent the wave-function consists
of an entangled superposition. 

We observe that the entropy rises sharply at $J \approx 0.25$~eV, 
corresponding to the transition from the doublet spin state to 
the triplet/quintet entangled state. 
As expected, the entropy is small in the low-spin region 
($J<0.25$~eV) and also in the fully-polarized limit. 
At $J=0$~eV (Fig.~\ref{fig2}.\textbf{b}), we find that the dominant configuration 
consists of the doublets (\dzd)$^2$(\dxyd)$^2$(\dxzd)$^2$,
with a single electron in the \dx orbital. 
The latter hybridizes strongly with the 
nitrogen $2p$ orbitals, but all other orbitals are mostly 
filled or empty, so this configuration is, essentially, 
a classical state with a finite magnetic moment.

At larger $J$ values, however, such as $J=0.8$eV (Fig.~\ref{fig2}.\textbf{c}),
all orbitals are partially filled, and an increasing 
number of electronic configurations, with different valence and spin, 
contribute to the statistics, and thus the iron impurity wave-function 
is \emph{fluctuating}.
Although the valence fluctuation are captured to some extent at the DFT level ($\Lambda_{DFT} \approx 0.75$),
we find that many body effects contribute significantly to the entropy.

Our results indicate that as FeP-d and FeP-p molecules approach 
a regime with large entanglement for $J\approx 0.5$, with a concomitant 
orbital selection close to the Fermi level. 
The orbital selection close to the Fermi level in turn induces a 
charge-localization effect. 
The latter effect of the Hund's coupling can be understood with
a simple picture:
a large Hund's coupling partially empties the \dz orbital and 
brings the weight of this orbital closer to the 
Fermi level, thereby reducing the hybridization 
between the iron $3d$ states and the 
nitrogen $2p$ states close to the Fermi level. 
The subtle interplay between the charge-localization induced by  
the Hund's coupling (orbital selection close to the Fermi energy) 
and the 
delocalization induced by strong correlations 
(the tendency for electrons to escape the iron
$3d$ orbitals in order to reduce the Coulomb energy)  
is captured by the DFT+DMFT 
methodology but is absent in Kohn-Sham DFT.
We emphasize that these ingredients are paramount 
to an estimation of the charge transfer
and binding properties between the 
iron atom and the ligand in oxyheme and carboxyheme. 


\begin{figure}
\begin{center}
\includegraphics[width=0.9\columnwidth]{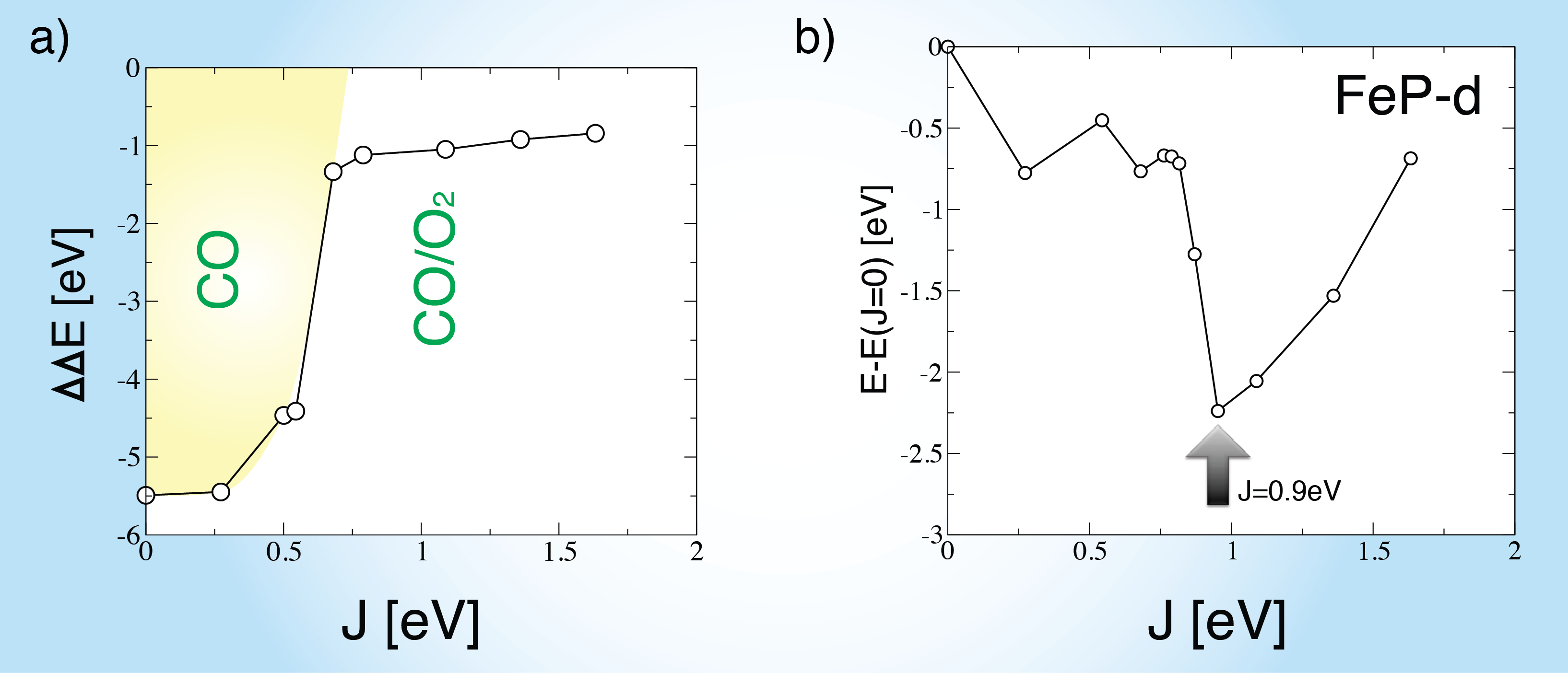}
\caption{ 
{\bf Energetics:} a) Difference in CO and O$_2$ binding energies $\Delta \Delta E$. The binding to CO is always
favored, however the imbalance is strongly reduced for $J>0.5$eV. b) Total energy of FeP as a function of $J$. The minimum
of the total energy is obtained for $J=0.9$eV.}
\label{fig4}
\end{center}
\end{figure}

\cred
Let us next \cblack discuss the effect of the Hund's coupling with respect to 
the unrealistic imbalance between the binding energies of CO and O$_2$ obtained by DFT. The binding
energy is defined as: $\Delta E =  E(FeP(X)) - \left(  E(FeP) + E(X)   \right)$, where X=CO or X=O$_2$.
The difference between the binding energies $\Delta E(CO) - \Delta E(O_2)$ is obtained by: 
$\Delta \Delta E = \Delta E_{CO} - \Delta E_{O_2}$.
For $J=0$eV, we find that the binding to CO is dramatically favoured,
when compared to the binding to O$_2$ (Fig.~\ref{fig4}.\textbf{a}): the difference in binding
energies is of the order of 5eV.
Although the binding to CO is favoured for all values of $J$, we
find that it is dramatically improved for $J>0.5$eV, and is reduced down to 1eV. This suggests that
other effects might be important to reduce further the CO/O$_2$ imbalance, such as that the effect
of the protein via the bending of the Fe-C-O angle \cite{cole}. 
 
 \cred
It is also worth noting that we find that the total energy of the molecule is minimized for $J=0.9$eV
(Fig.~\ref{fig4}.\textbf{b}), suggesting further that the heme molecule is particularly well suited to host
metallic d atoms, which tend to have a large screened $J$ interaction when hybridising to 
light elements such as nitrogen or oxygen. 
\cblack


\begin{figure}
\begin{center}
\includegraphics[width=0.65\columnwidth]{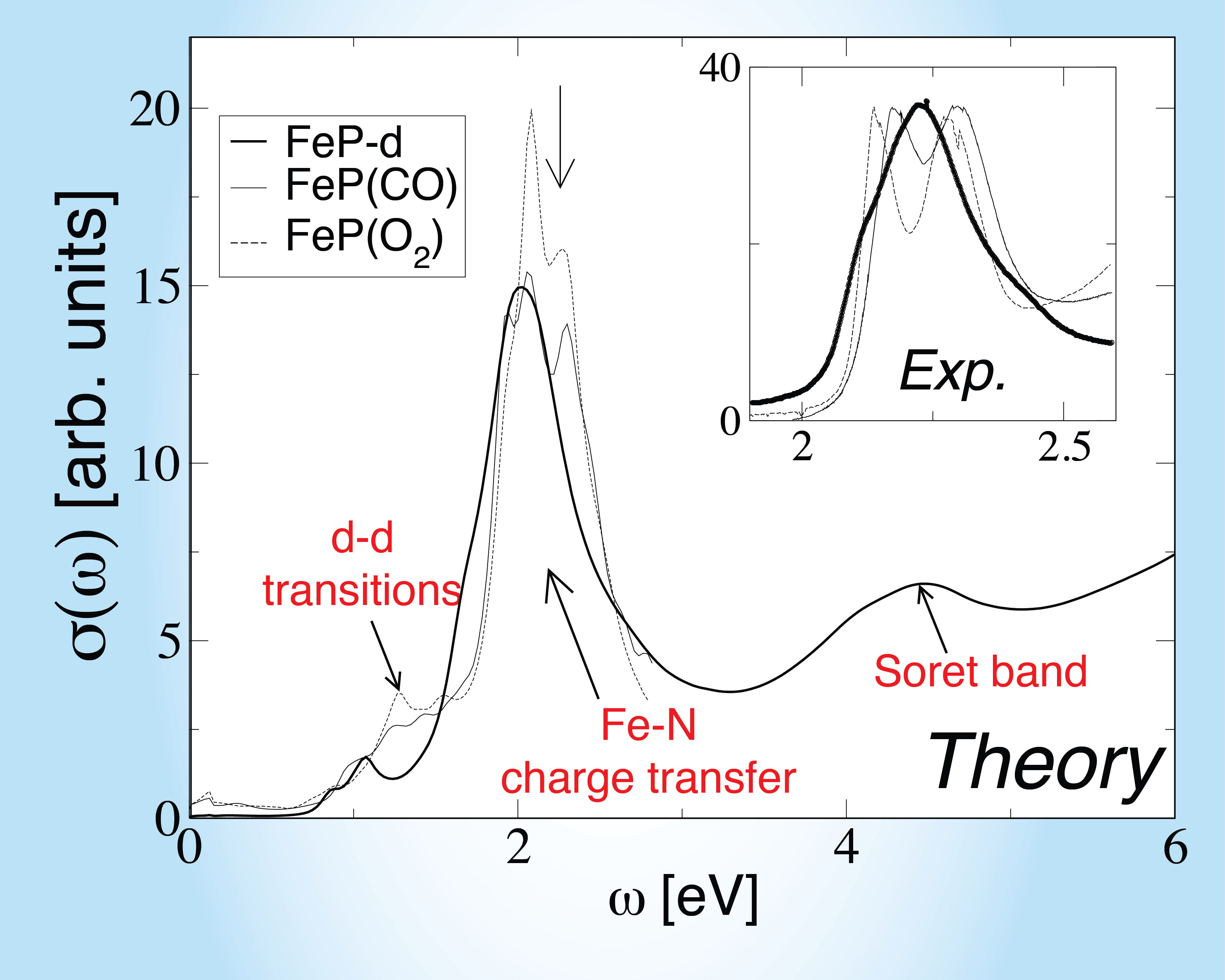}
\caption{ 
{\bf Optical measurements: } Optical conductivity of FeP-d (bold line) 
and FeP(CO) (solid) and FeP(O$_2$) (dashed line). The vertical arrow indicates
the energy of the experimental peak associated to the Fe-N charge transfer. Inset:
experimental measurements~\cite{heme_paper_optics} for unligated heme (bold line),
oxy- (dashed line) and carboxy- (solid) heme are shown for comparison.
$J=0.8$eV was used for all calculations above.}
\label{fig3}
\end{center}
\end{figure}
\cred We now \cblack move to our calculations of the 
optical absorption spectra of heme (Fig.~\ref{fig3}).
Our theoretical absorption spectra,
shown in Fig.~\ref{fig3}, are in \cred reasonable \cblack  
agreement with experimental 
data~\cite{heme_paper_optics}, in particular
for the optical transitions at $\omega \approx 2$~eV. 
We attribute this spectral feature to charge-transfer excitations 
from iron to nitrogen-centered orbitals.
The spectrum is dominated by the characteristic 
porphyrin Q-bands (those at $\approx 2$eV), and 
Soret bands~\cite{heme_soret_band} (at $\approx 4$eV).
Our results offer insight into 
the infrared absorption band 
present at $\approx 1$eV, in our calculation, and observed
in experiments at $0.6$~eV \cite{heme_infrared_optics}. 
This infrared peak is described, in our calculations,
as arising from transitions between 
the \dz spectral feature (HOMO) below the Fermi level 
and the LUMO (quasi-degenerate 
\dxz  and \dyzd) above the Fermi level.

Interestingly, we find that the infrared optical weight  
in unligated heme, associated with $d$-$d$ transitions and 
present in FeP-d, is absent in the planar theoretical model FeP-p.
Hence, the symmetry \cred breaking \cblack associated with the doming effect of the iron-intercalated 
porphyrin macrocycle \cred permits $d$-$d$ \cblack 
optical transitions, and is responsible for the spectral weight in the infrared regime.
We note that experimental spectra for FeP(CO) and FeP(O$_2$) exhibit a double peak structure at $\omega \approx 2$eV, absent
from our calculations done at $J=0$eV, but recovered for $J>0.8$eV. The best agreement with the experimental
data is obtained for $J=0.9$eV.
Finally, we extended our calculations to   
the time dependence of the magnetization of the iron atom 
after an initial quench in polarization (see sup. material). We 
propose that time-resolved spectroscopy 
may be used as a sensitive probe for the ligation state of heme.


In conclusion, we have carried out linear-scaling 
first-principles calculations, in combination with DMFT, on
both unligated and ligated heme. 
We have presented a newly-developed 
methodology applied to a molecule of important 
biological function, exemplifying how 
subtle quantum effects
can be captured by our methodology.
In particular, we have found that the Hund's coupling $J$
drives an orbital selection process
in unligated heme, which enhances the bonding
in the out-of-plane direction. 
The von Neumann entropy quantifying
valence fluctuations in the iron $3d$ subspace is
large for the physical values of $J \approx 0.8$eV.
This scenario sheds some light on the strong CO and O$_2$ binding
imbalance problem obtained by extracting the binding energies in 
simpler zero temperature and $J=0$eV DFT calculations. 
The difference in binding energies is dramatically reduced for
physical value of $J\approx 0.8$. The smaller remaining imbalance might be
further explained by the strain energy contained in the protein structure \cite{cole}
or by the contribution from the entropic term. Finally, the relevance of a finite Hund's coupling
in heme is confirmed by the total energy extracted from
the DFT+DMFT of unligated heme, which shows a minima for $J =0.9$eV. 

We have proposed a new 
mechanism for ligand binding to heme based
on an orbital selective process, on this basis,
a scenario which we term 
\emphasize{bonding determined by local 
valence fluctuations}.
Finally, we have obtained a reasonable agreement 
between experimental and our theoretical 
optical absorption spectra, our description accounting 
for the observation of optical transitions in the 
infrared regime
and the double peaked structure of the optical response at $\omega \approx 2$eV.

At the time of writing, we became aware of related application of DMFT to 
\cred an organometallic crystal \cblack 
\cite{roser_valenti_biology_dmft}.
We are grateful to R.H.~McKenzie for comments and bringing Ref. \onlinecite{porphyrin_mixed_state} to our attention, and to
D.~Cole for many insightful discussions.
C.W. was supported by the Swiss National 
Foundation for Science (SNFS).
D.D.O'R. was supported by EPSRC.
N.D.M.H was supported by EPSRC grant number EP/G055882/1.
P.B.L is supported by the US Department of 
Energy under FWP 70069. 
Calculations were performed on the Cambridge High Performance Computing Service under 
EPSRC grant EP/F032773/1. Correspondence 
and requests for materials should be addressed to C.W.



%

\end{document}


\title{Importance of many body effects in the kernel of hemoglobin for ligand binding}

\author{C\'edric Weber}
\affiliation{King's College London, Theory and Simulation of Condensed Matter (TSCM), The Strand, London WC2R 2LS}
\affiliation{Cavendish Laboratory,  J. J. Thomson Av., Cambridge CB3 0HE, U.K.}
\author{David D. O'Regan}
\affiliation{Cavendish Laboratory,  J. J. Thomson Av., Cambridge CB3 0HE, U.K.}
\affiliation{Theory and Simulation of Materials, \'{E}cole Polytechnique F\'{e}d\'{e}rale de Lausanne, Station 12, 1015 Lausanne, Switzerland}
\author{Nicholas D. M. Hine}
\affiliation{Cavendish Laboratory,  J. J. Thomson Av., Cambridge CB3 0HE, U.K.}
\affiliation{Department of Materials, Imperial College London, Exhibition Road, London SW7 2AZ, U.K.}
\author{Peter B. Littlewood}
\affiliation{Cavendish Laboratory,  J. J. Thomson Av., Cambridge CB3 0HE, U.K.}
\affiliation{Physical Sciences and Engineering, Argonne National Laboratory, Argonne, Illinois 60439, U.S.A.}
\author{Gabriel Kotliar}
\affiliation{Rutgers University, 136 Frelinghuysen Road, Piscataway, NJ, U.S.A.}
\author{Mike C. Payne}
\affiliation{Cavendish Laboratory,  J. J. Thomson Av., Cambridge CB3 0HE, U.K.}


\begin{abstract}
In this supplemental material,
we study the dynamical magnetic response of 
heme, in various ligation and doming configurations,
by means of out-of-equilibrium   
calculations performed by perturbing the 
DFT+DMFT ground state wave-function and propagating in the
Keldysh time domain. 
Based on the rather strong ligand specificity observed in these 
simulations, we propose a potentially sensitive method for
profiling of heme binding, 
using two-photon circularly-polarized spectroscopy.
Here, we also describe the theoretical methods 
used in the work presented in our Letter in detail, and justify the 
range of parameters used in our study, i.e. we explore the dependence
of our calculations on the choice of the Coulomb repulsion and the temperature.
A simple toy model for calculating the entropy is proposed, and compared to the DMFT calculations, in order
to illustrate the importance of many body effects.
We furthermore justify the importance of the Hund's coupling based on phenomenological grounds,
by carryout a comparison between the theoretical optics at various values of the Hund's coupling with experiments. 
Out-of-equilibrium calculations are also discussed.
\end{abstract}

\maketitle


In this work, we have carried out a detailed theoretical study
 of the electronic structure of the heme molecule by 
means of a  combination~\cite{our_paper_vo2,oregan12}
of linear-scaling density-functional theory (DFT) 
with the dynamical mean-field theory 
approximation (DFT+DMFT)~\cite{OLD_GABIS_REVIEW,
dca_reference_cluster_dmft},
a model which includes local dynamical, finite-temperature 
and multi-determinental effects, for given Hamiltonian
parameters, at an effectively exact level.
Although DFT+DMFT has been primarily applied to solids to date, 
here we apply our real-space DFT+DMFT implementation
to a substantial molecular system, extending the scope of 
DFT+DMFT applicability to biology for the first time to our
knowledge.
We performed a study of the impact of varying the 
Hund's exchange coupling strength $J$.
Although the value of J in iron based molecules and solids is expected
to be $J \approx 0.8$~eV,~\cite{hund_value_iron_arsenide,iron_arsenide}
we are interested in this work to study what are the physical ingredients introduced by
J as it is increased from zero.
In particular, it was suggested that the Hund's coupling might fix the strength of the correlations~\cite{luca_prl_hund_janus},
or act to increase 
the total magnetic moments within the localized 
effective impurity subspaces 
treated using DMFT and, in doing so, 
break degeneracies between the localized orbitals.

The effects on the electronic structure due to the diatomic axial ligands 
O$_2$ and CO are considered, and testing is further carried
out with respect to changes in the Hubbard 
$U$ parameter and temperature.
Our study furthermore includes two experimental signatures
of ligand binding, namely the optical conductivity 
discussed in our Letter,
and the transient magnetic response with which we discuss 
at the end of this supplemental material.

\section{Methodology: Density-functional theory 
calculations using optimized Wannier functions}
 
Kohn-Sham density-functional theory 
(DFT)~\cite{dft_kohn, kohn_sham} was used in this
study to provide a reasonable, conveniently-computed 
zero-temperature, single-determinental
 starting point for the
DFT+DMFT self-consistency procedure, within the
localized subspace treated for many-body effects using DMFT,
as well as a sufficiently 
detailed description of the delocalized electron
``bath" comprising the remainder of the system.
Relatively simple local and semi-local approximate functionals
for DFT often provide an acceptably refined 
description of the ground-state electronic structure, and 
even the quasiparticle band-structure, associated with orbitals of large
spread or no magnetic order. This observation  
forms the basis of the DFT+DMFT
approximation, in which the dominant computational effort is instead 
focused on solving for the many-body wave-function 
constructed from the 
the remaining, localized, possibly
spin-polarized orbitals adjacent to the Fermi-level.

The ONETEP 
linear-scaling DFT and DFT+$U$ 
code~\cite{onetep,linearscalingdftu,onetep_ref6} 
 was used
to solve for the DFT ground-state subspace Green's function
in this work.
The ONETEP method is based on direct minimization of
the total-energy with respect to the single-particle density-matrix,
and is particularly advanced in terms of its accuracy 
equivalent to that of a plane-wave method,
which is arrived at by means of an \emph{in situ} 
variational optimization of the  
expansion coefficients of a minimal set of spatially-truncated 
Nonorthogonal Generalized Wannier 
Functions~\cite{onetep_ref_ngwf} (NGWFs)
$ \left\lbrace \phi_\alpha \right\rbrace$ (with non-trivial overlap
$S_{\alpha \beta} = \langle \phi_\alpha \lvert
\phi_\beta \rangle $).
The basis for this expansion is a set of systematically improvable 
Fourier-Lagrange, or psinc functions~\cite{onetep_optimisation_ngwfs}, which is 
equivalent to a truncated set of plane-waves.
We express
the single-particle density-matrix in the separable form~\cite{vo2_paper_ref_advance_in_dft_separable_form, linear_scaling_separable_form_density_onetep} given by
\begin{equation}
\rho_0 \left( \mathbf{r}, \mathbf{r'} \right)
=  \sum\limits_{\alpha \beta} 
{\phi_\alpha \left(  \mathbf{r} \right) K^{\alpha \beta} \phi_\beta
\left(  \mathbf{r'} \right)},
\end{equation}
where  $K^{\alpha \beta} = \langle 
 \phi^\alpha | \hat{\rho}_0 |\phi^\beta \rangle $ is 
known as the density-kernel, the 
tensor representation of the single-particle density operator $\hat{\rho}_0$~\footnote{Repeated 
index  pairs are implicitly summed over, 
as per the Einstein summation convention; 
enclosure by parenthesis suspends the summation.}.
The contravariant NGWF duals $\left\lbrace \phi^\alpha
\right\rbrace $ are defined as those which satisfy the
biorthonormality relation, where the overlap matrix acts as metric tensor,
defined by 
\begin{align}
\langle \phi^\alpha \lvert
\phi_\beta \rangle= \delta^{\alpha}_{ \beta }, \quad \mbox{with}
\quad \lvert \phi^\alpha \rangle = \left( S^{-1} \right)^{\alpha \beta} 
 \lvert \phi_\beta \rangle.
\end{align}
The use of a minimal, optimized Wannier function 
representation of the density-matrix
allows for the DFT ground state to be solved with relative ease in large systems, 
particularly in molecules where their explicit truncation implies that the
addition of vacuum does not increase the computational cost.
Furthermore, it allows for the full Kohn-Sham Green's function
to be computed via Hamiltonian inversion with
a manageable computational overhead. 
Thus, we do not make 
any projection over bands or orbitals prior to generating 
the full Green's function, and only project 
after inversion when computing the matrix 
elements of the subspace Green's function.

\section{Methodology: Ionic geometries, 
pseudopotentials and variational parameters}

We performed ground-state  
DFT calculations on heme 
\emph{in vacuo} by iteratively minimizing the energy functional 
with respect to both the density kernel and the 
NGWFs, using the algorithm  
described in Ref.~\onlinecite{onetep_energy_min}.
The total-energy
was converged to within
$1$~meV per atom, with respect to the 
plane-wave energy cutoff 
(at a minimum of $1300$~eV for all reciprocal lattice vectors in our calculations) 
and the NGWF cutoff radii 
(at $6.6$~\AA~ for all species), 
and no additional restrictions on the variational freedom, such as the density 
kernel truncation, were invoked.
The simulation cell volume was increased until
(at $4.8 \times 10^4$~\AA$^3$)
periodic-image interaction errors also fell within
this total-energy tolerance.

The DFT calculations were performed, and pseudopotentials
generated, using the PBE~\cite{PhysRevLett.77.3865} 
generalized gradient exchange-correlation functional. 
In line with common practice in DFT+DMFT calculations, 
in which all magnetic order 
is assumed to be confined to the localized impurity subspaces,
spin-degenerate DFT ground-state densities were used. 
Scalar scalar-relativistically corrected 
norm-conserving PBE~\cite{PhysRevLett.77.3865}  pseudopotentials, 
generated with the OPIUM package~\cite{opium_package},
were used in place of the core electrons for all ions.
The semi-core states of iron were included in the valence, so that the 
(3s,3p,3d,4s,4p) sub-shells were were retained for 
explicit treatment,
obviating the use of a non-linear core correction.
For the lighter element, (2s,2p) were retained in the valence
for C, N, and O, along with (1s) for H.
Hence, we used 13 variationally-optimized NGWFs to 
describe the Fe electrons, 4 NGWFs for each C, N, and O atom, 
and 1 for each H atom, 
giving a total of 213 NGWFs (spin-degenerate orbital pairs) 
for the unligated heme system
and and 221 each for the two ligated systems.

Ionic geometries were obtained from the 
Protein Data Bank:  unligated  
deoxyheme (dome-shape, FeP-d) was extracted from a
human deoxyhaemoglobin  
structure obtained by x-ray 
crystallography (PDB key 2HHB \cite{2HHB_pdb_iron_porphyrin}), and the
missing hydrogens were
reintroduced using the Jmol package \cite{jmol}. 
The Carboxyheme (heme-CO complex, FeP(CO)) and
the oxyheme (heme-O$_2$ complex, FeP(O$_2$)) structures were
extracted using the same procedure 
(PDB key 1GCW \cite{1GCW_pdb_porphyrin} and 1HHO \cite{1HH0_pdb_iron_porphyrin},
 respectively). 
A theoretical, planar deoxyheme (FeP-p) (PDB key HEM) is 
also considered in our calculations for 
comparison with the experimentally-observed dome-shaped 
structure in order to test how the conformal 
change between the planar and 
dome-shape affects the electronic structure.
In the present work, we limit our 
calculations to a single heme molecule,
omitting residues which neighbour it in the
haemoglobin protein.

\section{Methodology: Definition of the low-energy, correlated model}
 
We extended our DFT calculations using the DFT+DMFT 
method~\cite{OLD_GABIS_REVIEW,dca_reference_cluster_dmft} in order to 
refine the description of strong correlation effects arising due to partial
degeneracies  between the localized orbitals making up the
iron $3d$ sub-shell at the core of heme.
DMFT allows for quantum and thermal fluctuations, absent 
from the zero-temperature, single-determinant Kohn-Sham DFT calculations,
to be accurately described.
The heme molecule was represented within DMFT by an 
Anderson impurity Hamiltonian (AIM)~\cite{heme_aim_kondo}, 
in its Slater-Kanamori form,
the ground-state of which was solved 
for by using a finite-temperature Lanczos solver \cite{cpt_ed_solver}.
Since only a single impurity site ($3d$ orbital subspace)
is present, the system becomes crystal momentum independent
in the molecular limit, and since the Kohn-Sham Green's 
function is inverted in full before projection 
onto the impurity subspace, 
the Anderson impurity mapping is effectively exact, 
so that the DFT+DMFT algorithm
is almost entirely converged in a single 
Green's function self-consistency step.
We note that a Kondo model was used early in the literature to
study heme, along similar lines than ours, by means of a model Hamiltonian \cite{heme_aim_kondo}.
In our study, the AIM model is determined and obtained by first-principle calculations.

A spherically symmetric trial 
impurity subspace was
defined by a spanning set of iron $3d$
orbitals, the orthonormal set $\left\lbrace \varphi_m \right\rbrace$ 
produced by solving the spherically-symmetric Schr\"odinger equation
subject to the norm-conserving 
iron pseudopotential with an appropriate confining potential, 
the within a truncation sphere of radius $6.6$~\AA. The same procedure
was used to generate the initial guesses for the NGWFs
during the initialization of the DFT calculation, so that the 
trial impurity subspace formed 
a proper subspace of the initial guess for the
Kohn-Sham Hilbert space. 
Since the latter is optimized as the energy is 
minimized with respect to the NGWFs, so, at convergence
of the DFT algorithm, the Hubbard
projectors finally chosen to span the 
impurity subspace of DFT+DMFT were the so-called symmetry-adapted 
Nonorthogonal Generalized Wannier Functions (SNGWFs), 
defined by
\begin{align}
\lvert \tilde{\varphi}_m \rangle = \lvert \phi_\alpha \rangle
\langle \phi^\alpha \rvert \varphi_m \rangle 
= \lvert \phi_\alpha \rangle \left( S^{-1} \right)^{\alpha \beta}
\langle \phi_\beta \rvert \varphi_m \rangle.
\end{align}
Thus, the final impurity subspace is a proper subspace of the
converged Kohn-Sham Hilbert space, 
a necessary condition for a strictly causal self-energy, 
but retains the $3d$ symmetry
of the numerical atomic orbitals (the generalization to 
nonorthogonal projector functions in this context
is discussed extensively in Ref~\onlinecite{dftu_david_subspace_representation}).

Once the fully-converged DFT energy minimization of was carried out, 
for each heme strain, the full 
Green's function was initially computed in the finite-temperature Matsubara 
representation.
Noting that 
the inverse of a doubly-covariant tensor is a doubly-contravariant
tensor, the full Green's function is generally expressed
and computed, in terms of the
the Kohn-Sham Hamiltonian $\bold{H}$, as
\begin{equation}
\label{fullg}
G^{\alpha \beta} \left(i \omega_n \right) = \left( ( i \omega_n+\mu )S_{\alpha \beta}  - H_{\alpha \beta} -  \Sigma_{\alpha \beta}  \right)^{-1}.
\end{equation} 
Here, $\mu$ is the chemical potential,
set to the average of the highest occupied and
lowest unoccupied Kohn-Sham orbital energies,
and $\bold{\Sigma}$ is the 
self-energy tensor generated by the DMFT algorithm, where 
$\bold{\Sigma} = \bold{0}$
in the first instance so that the initial full Green's function is the
Kohn-Sham Green's function $\bold{G}_{0}$. 
Green's function sampling at $400$ 
Matsubara frequencies provided adequate convergence of
the properties of interest.
We performed this matrix inversion, as well as all 
matrix multiplications involved in the DMFT algorithm,
on graphical computational units (GPUs) using a tailor-made parallel implementation of the LU decomposition 
using the CUDA programming language. This provided a crucial improvement in
the computational feasabililty of our calculations.
Following inversion to find the the full 
Green's function, the Kohn-Sham subspace Green's 
function $\bold{\tilde{G}}_0$ is given by its projection onto the
impurity subspace, where it has the matrix representation 
\begin{align}
\label{projgreen}
{\tilde G}_{ 0 m m'} \left(i \omega_n \right) &{}= 
\langle \tilde{\varphi}_m \rvert 
\hat{G} \left(i \omega_n \right) 
\lvert \tilde{\varphi}_{m'}
\rangle \\
&{}= 
W_{ m \alpha}   G^{\alpha \beta}  \left(i \omega_n \right) V_{ \beta m'}, \nonumber
\end{align}
where $m$ and $m'$ run over the five iron 
$3d$ SNGWF projector functions
(in real cubic-harmonic notation: \dxd, \dzd, \dyzd, \dxzd, \dxyd), $\alpha$ and $\beta$ are the indices for the NGWFs,
and the matrices NGWF-projector overlap matrices 
are defined as ${V}^{\left(I\right) }_{\alpha m} = 
\langle  \phi_\alpha | \varphi^{\left( I \right)}_m \rangle$ and ${W}^{\left(I\right) }_{m \alpha} = 
\langle  \varphi^{\left( I \right)}_m |  \phi_\alpha \rangle$.

In practice, in order to imbue the SNGWF Hubbard projectors with a more plausible
physical interpretation,  a real-space rotation of the 
functions was carried out~\cite{rotation_harmonics} in order to better align their lobes.
The subspace projected Green's function is thus transformed to 
\begin{equation}
\bold{\tilde G^{rot}}=   \bold{\tilde U}^\dagger  \bold{\tilde G} \bold{\tilde U} ,
\end{equation}
where $\bold{\tilde U}$ is the $5\times5$ rotation matrix in cubic harmonic space, corresponding to a  rotation in $R^{(3)}$, and 
$\bold{\tilde G^{rot}}$ is the rotated subspace Green's function passed to the DMFT solver. 
The rotation matrix $\bold{\tilde U}$ is chosen, pragmatically, 
such that the $\bf{e}_x$ and $\bf{e}_y$ axes are those, in an averaged sense,
which point towards the four nitrogen atoms surrounding the iron-centered
impurity subspace, with $\bf{e}_z$ directed out of the porphyrin plane. 

Electronic correlation effects beyond the capacity of the approximate
exchange-correlation functional, 
those arising due to interactions between particles within the 
impurity subspace and finite-temperature effects, 
are explicitly described in our DMFT calculations by the 
Slater-Kanamori form of the Anderson impurity Hamiltonian~\cite{slater_kanamori_interaction,hund_coupling_kanamori},
specifically
\begin{align}
\label{hint}
\mathcal{H}_{U} ={}& U \sum \limits_{m} {n_{m \up} n_{m \dn}} + (U' -\frac{J}{2})\sum\limits_{m>m'}{n_{m}n_{m'}}  \\
\nonumber {}&
-J \sum\limits_{m>m'}{\left( 2 \bold{S}_m \bold{S}_{m'} + \left(  d^\dagger_{m\up} d^\dagger_{m \dn } d_{m' \up} d_{m' \dn}   \right) \right)    }.
\end{align}
In this, the first term describes the effect of 
intra-orbital Coulomb repulsion, parametrised by $U$, and the second
term describes the inter-orbital repulsion, proportional to $U'$,
which is renormalized by the Hund's exchange coupling 
parameter $J$ in order
to ensure a fully rotationally invariant Hamiltonian (for 
further information on this topic, we
refer the reader to Ref.~\onlinecite{imada_mott_review}). 
The third term is the Hund's rule exchange coupling, 
described by a spin exchange coupling of amplitude $J$.
$\mathbf{S}_{m}$ denotes the spin corresponding to orbital $m$,
so that $S_{m}=\frac{1}{2}d_{m s}^{\dagger}
\vec{\sigma}_{s s'}
d_{m s' }$, where $\vec{\sigma}$ is the vector of Pauli matrices
indexed by $s$ and $s'$.
We note that the Slater-Kanamori form of the vertex interaction is especially
well suited to capture the multiplet properties of the low energy states \cite{imada_mott_review}.
The Slater-Koster \cite{Slater54} interaction is another alternative, but is mostly used to describe solids, and is not well
suited to the case of a molecule.

In this this work, unless where otherwise stated,
we used $U=4$~eV for the screened 
Coulomb interaction~\onlinecite{heme_marzari} and
we explored the dependence of several observables on the HundÕs coupling (in the range $J=0-2.5$~eV).
Our DMFT calculations were carried out at room 
temperature, $T=294$~K, again except where otherwise stated.
The dependence of our results with respect to the Coulomb repulsion $U$ and the temperature $T$
is also discussed in the next sections.

\section{Methodology: Dynamical mean-field theory treatment}

The Hamiltonian (\ref{hint}), in combination with the
expressions 
(\ref{fullg}) and (\ref{projgreen}), defines the many-body 
impurity problem that we 
solve using the DMFT algorithm~\cite{OLD_GABIS_REVIEW}, 
which updates the impurity Green's function $\bold{\tilde G}$.
The DMFT establishes a self consistent mapping from the correlated atoms
of the initial solid or molecule and a smaller local problem, the Anderson Impurity Model (AIM),
which is used to obtain the self energy within this projected subspace ~\cite{OLD_GABIS_REVIEW}.
The mapping is carried out self consistently, and the obtained local Green's function of the AIM
converges to the Green's function of the larger space projected onto the correlated atom.
At the level of the AIM, this model describes an impurity connected to a bath by the so-called
hybridization function. The bath of the AIM models the surrounding environment of the correlated atom in the
solid or the molecule, and the hybridization function describes how electrons are dynamically transferred from and to
the bath to the impurity. Hence, at the AIM level, the uncorrelated part of the Heme molecule is described by the bath,
and the hybridization between the d orbitals of the iron atom and the N atoms is described by the hybridization function in the AIM. 

We define only a single impurity subspace in 
calculations on heme, since there is only one transition-metal
ion present, and so the impurity self-energy 
\begin{align}
\bold{\tilde \Sigma} = \bold{\tilde G}_0^{-1} - \bold{\tilde G}^{-1}
\end{align} 
resulting from the DMFT algorithm is said to be exactly \emphasize{local}, 
in the sense that there are no pairs of impurity sites 
for which to consider site off-diagonal self-energy matrix elements.
In this particular case, there is no feedback from the self energy to the hybridization function, and
the DMFT converges after one iteration. It is hence not a mean-field approximation in this particular case,
but turns out to be exact. However, the methodology described in this work can be applied to molecules
with many correlated ions with no further modifications. 

The hybridization matrix $\bold{\Delta}$ within the AIM impurity subspace
is given, formally, by 
\begin{equation}
\label{hybrid}
\bold{\Delta}(i\omega_n) = \left( i \omega_n+ \mu \right) 
\bold{\tilde O}  - \bold{\tilde \Sigma} - \bold{E^{imp}} - \bold{\tilde G}^{-1},
\end{equation}
where here~\cite{non_orthogonal_dmft,non_orthogonal_dmft_kotliar,
non_orthogonal_basis_cdmft}, 
the metric tensor on the SNGWFs is given by
\begin{equation}
\bold{\tilde O} =
\left( 
\bold{W} 
\bold{ S}^{-1} 
\bold{V}
\right)^{-1}.
\end{equation}
This metric is in general non-trivial, i.e., $\bold{\tilde O} \ne \bold{1}$
and so the SNGWFs are nonorthogonal and not identical to their duals,
even if their parent atomic orbitals form an orthonormal set,
if the trial impurity subspace does not form a proper subspace of the
converged Kohn-Sham Hilbert space.
However, for this particular case of study, we found however that the deviation of $\bold{\tilde O}$ 
from the identity matrix was small (within $0.1\%$).

The high-frequency part of the hybridization 
function, $\bold{E}^{imp}=\bold{\Delta} ( i \omega_n \rightarrow \infty)$, defines the impurity energy levels
in the zero-hybridization ``atomic" limit, so 
that the hybridization matrix $\bold{\Delta}(i\omega_n)$ 
exhibits the correct physical decay proportional to $ 1/ i\omega_n$.
We tested that the implied limiting condition $\bold{\tilde O}= 
\lim_{  \omega \rightarrow \infty} 
\left[  {\bold{\tilde G}}^{-1}  \left( i \omega \right)''/\omega \right] $
holds, up to a high precision, ensuring that the self-energy is physically meaningful.
It can also be straightforwardly obtained by doing a high frequency expansion of the Green's function
that:
\begin{equation}
E^{imp} = \bold{\tilde O}  \bold{W} \left( \bold{S}^{-1} \bold H \bold{S}^{-1} \right)  \bold{V}  \bold{\tilde O}.
\end{equation}

The self-energy 
$\bold{\tilde \Sigma}$ 
is thus obtained by solving the Anderson impurity 
model (AIM) defined by the hybridization (\ref{hybrid}) 
and the interaction Hamiltonian (\ref{hint}),
using  a finite-temperature Lanczos DMFT algorithm~\cite{medici_finite_temperature_lanczos,
finite_temperature_lanczos_leibsch,our_plaquette_paper}.
The Lanczos solver uses a finite 
discretization of the hybridization (\ref{hybrid}),
suffering finite size effects, 
yet allows for the orbital off-diagonal elements of
the hybridization (\ref{hybrid}) to be considered on an 
equal footing to the diagonal elements.
Crucially, this mitigates some of the dependence on the
spatial form and orientation of the projector functions (SNGWFs)
used to define the impurity subspace.

Having solved for impurity self-energy 
given by the AIM, $\bold{\tilde \Sigma^{rot}}$, 
it is rotated back to
the original system of coordinates, to give
\begin{equation}
\bold{\tilde \Sigma}=  \bold{\tilde U} \bold{\tilde \Sigma^{rot}} 
\bold{\tilde U}^\dagger.
\end{equation}
We then ``up-fold" it to the Kohn-Sham
Hilbert space, using
its separable form in terms of NGWFs, that is
 \begin{equation}
\label{sigmadc}
  {\Sigma}_{\alpha \beta} =
  V_{ \alpha m}  
\left( \tilde{\Sigma}^{m m'} 
- V_{\mathrm{dc}}  \tilde{O}^{m m'} 
\right)
 W_{m' \beta}.
 \end{equation} 
 A separable form of the up-folded self-energy enforces
 its causality, as shown recently in 
 Ref~\onlinecite{dmft_lda_implementation}, 
(the local self-energy for the impurity subspace being causal 
by construction, so that ${\tilde \Sigma_{m m'} (i \omega_n)}'' \le 0, \;\;
 \mbox{for all} \;\; i\omega_n, m, m'  $.
We carefully verified that our computed self-energies were causal, 
a necessary condition for positive definite spectral functions and
the physicality of computed observables generally.
In this work, we used the canonical 
form of the double-counting potential $V_{\mathrm{dc}}$, 
given by
\begin{equation}
\label{dc}
V_{\mathrm{dc}}= U^{av} \left( n_d - \frac{1}{2} \right) - 
\frac{J}{2} \left( n_d - 1 \right),
\end{equation}
where $n_d$ is the invariant total occupancy 
of the impurity subspace, 
\begin{align}
n_d = \frac{2}{\pi} \int_0^\mu d\omega \; \tilde{G}_{m m'} \left( \omega \right) \tilde{O}^{m' m}.
\end{align}
Here, the parameter  $U^{av}$ 
is the averaged repulsion, related 
to the intra-orbital and inter-orbital repulsion, 
and computed using the expression~\cite{hund_coupling_averaged_double_counting}
\begin{equation}
\label{uav}
U^{av}= \frac{U+ 2 \left( N_{\mathrm{deg}} -1 \right) U'}{2N_{\mathrm{deg}}-1},
\end{equation}
where $N_{\mathrm{deg}}=5$ is  the number of localised 
orbitals (SNGWFs) spanning the
impurity subspace.
The double-counting correction is intended to subtract the interactions
within in the impurity subspace that are already included, to some extent, 
in the Kohn-Sham Hamiltonian $\bold{H}$.
This is carried out in an approximate, averaged fashion, since the exact
form of double-counting correction required for a given density, exchange-correlation
functional, and set of projecting orbitals, is not known.
A good test of the validity of the double-counting correction, verified in
our calculations, is the independence of the impurity 
chemical potential 
with respect to the Coulomb interaction parameter $U$. 
The equations (\ref{fullg}), (\ref{projgreen}), (\ref{hybrid}), and (\ref{sigmadc}) 
together form the Green's function self-consistency cycle used in our DFT+DMFT
calculations, which is iterated until convergence is obtained. In the calculations
presented in this work, for reasons aforementioned, the mapping onto the AIM is
effectively exact, and the DFT+DMFT self-consistency cycle converges rapidly, 
being already close stationary after a single iteration.

\section{Methodology: iron density and magnetization, entanglement and entropy}
 
Once the AIM problem is solved with the finite temperature Lanczos algorithm, we have access to the ground state
and the low energy excited states.  
The eigenstates of the many-body Hamiltonian are written in the Fock space of the impurity connected to the bath.
The basis states of the Fock space span all the possible electronic configurations, hence an eigenstate
of the hamiltonian is a superposition of all possible quantum states of the impurity and the bath. 
It is important to note that other simpler approaches, such as the so-called \emphasize{Configuration interaction} \cite{configuration_interaction_CI}
only consider a very restricted basis set where a few electrons are excited from the impurity to the bath. 
In our approach, this is not the case, and all possible configurations are taken into account.  

Observables related to the impurity model, such as the impurity density, can be obtained
by the finite temperature thermal average. The trace of an operator $\hat A$ is defined as:
\begin{equation}
 \langle \hat A \rangle  = \frac{\sum\limits_{i}{e^{-\beta E_i } \langle i | \hat A | i \rangle}}{Z},
\end{equation}
where $Z$ is the partition function, $\beta=1/T$ is the inverse temperature, and the index $i$ runs over the low energy states. 
The impurity density operator $n_d$ is simply defined as:
\begin{equation}
 \hat n_d = \sum\limits_{\sigma}c^\dagger_{i\sigma} c_{i\sigma}
\end{equation}
where $c^\dagger_{i\sigma}$ ($c_{i\sigma}$) creates (destroys) an electron in the orbital $i$ of the iron d manifold with spin $\sigma$. 
The thermal average of $n_d$ is given by:
\begin{equation}
  \langle n_d \rangle = \frac{ \sum\limits_{i,\alpha,\sigma}{ e^{-\beta E_i} \langle i | c^\dagger_{\alpha\sigma} c_{\alpha\sigma} | i \rangle}}{Z},
\end{equation}
where $\alpha$ denotes the 5 d orbitals, $i$ runs over the eigenstates of the AIM, which is 
determined self-consistently by the DMFT, $E_i$ is the eigen energy of the quantum eigen state $ | i \rangle$.
We note that the summation over the eigenstates $| i \rangle$ involves states with different number of electrons, and hence
$n_d$ is not an integer number, but can take any rational values.
Since upon DMFT convergence the impurity Green's function converges to the Green's function of original problem,
the density $n_d$ of the impurity model corresponds to the electronic occupation of the iron atom in heme.

We note that, although the total number of electrons, and 
global observables
more generally, are associated with operators that 
commute with the Hamiltonian and, hence, are 
conserved quantities, 
the density at the iron site is a local observable 
and can fluctuate strongly.

In particular, we find for $J=0$~eV that DFT+DMFT 
yields an impurity occupancy of $6.8$~e for deoxyheme,
(both for FeP-p and FeP-d), 
with larger subspace occupancies for ligated hemes ($n_d=7.0$~e for
FeP(CO) and $n_d=6.9$~e for FeP(O$_2$)). 
This is in good agreement with earlier DFT+$U=6$~eV  
 calculations~\cite{david_thesis},
which gave $n_d=6.6$~e for deoxyheme and $n_d=6.8$~e for carboxyheme. 
We note that doming of the heme molecule 
has no effect on the subspace occupancy at small Hund's coupling,
although a significant dependence of the 
occupancy on the doming is activated beyond
$J \approx 0.3$~eV.

In our calculations we found that the DMFT solution remains paramagnetic, although it was allowed for the possibility
to spontaneously form a magnetic moment (spin symmetry broken state). However, the low energy states are in a quantum
superposition of polarized states, giving a fluctuating magnetic moment at the iron site, which can be quantified as \cite{spin_correlation_iron}:
\begin{equation}
S = \sqrt{ \langle \hat {\bold{S}} \hat {\bold{S}} \rangle - (\langle \hat {\bold S} \rangle)^2  }=\sqrt{ \langle \hat{\bold{S}} \hat{\bold{S}} \rangle }
\end{equation}
Where $S$ is the obtained magnetic moment of the iron atom, $\bold S$ the total spin operator at the iron site. The spin $s$ is obtained as usual by the relation:
\begin{equation}
S = s  \left( s + 1 \right).
\end{equation}

We note that the eigenstates obtained by the Lanczos are written in the Fock space which contains
both the impurity orbitals and the bath orbitals, and although the iron density $\hat n_d$ operator is
defined in terms of d orbitals, the low energy states are defined in the basis of both the impurity and the
bath orbitals. The entanglement between the impurity and the bath can be quantified by the reduced density matrix of the impurity $\hat \rho$,
defined as: 
\begin{equation}
  \hat \rho = \sum\limits_{i} e^{-\beta E_i} Tr_B | i \rangle \langle  i |, 
\end{equation}
where $Tr_{B}$ denotes the partial trace over the bath degrees of freedom, and $|i\rangle$ are the low energy states of the problem.

In particular, the eigenvectors $\bold v_k$ of $\hat \rho$ provide a simplified description of the many body states spanned by the 
impurity: they give a cartoon picture in terms of electronic orbitals for the impurity at low energy.
The eigenvalues of $\hat \rho$, $\lambda_k$ are normalized weights ($\sum\limits_{k} \lambda_k=1$)
which give the probability to measure the corresponding many body state. We note that:
\begin{itemize}
\item 
 each $\bold{v}_k$ is a superposition of many body states of the Fock space, so it is a quantum superposition of electronic occupations of the
 impurity orbitals
\item
 there is in general not only one dominant $\lambda_k$, but a collection of eigenvalues with similar amplitudes.
\end{itemize}
The Von Neuman entropy, defined as:
\begin{equation}
 \Lambda = - k_B \sum \limits_k { \lambda_k \text{ln}  \left( \lambda_k \right) }
\end{equation}
measures the entanglement between the impurity and the bath. When a number of $\lambda_k$ with similar amplitudes starts
to proliferate, the entropy becomes large. 
In the particular case when $\lambda_1=1$, the impurity is in a pure state and the associated entropy $\Lambda=0$. In the latter case,
the quantum state of the impurity is well defined, and a spin and a charge can be used to label the ground state of impurity problem. 

As discussed in the last section, since in the case of DMFT applied to heme there is only one correlated site, the impurity observables,
such as the impurity density $n_d$ or the entropy $\Lambda$, readily provide respectively the iron charge density and the entropy associated
with the iron atom in the heme molecule. Hence, if the impurity in the AIM is in a pure state ($\Lambda=0$), one might associate to the heme molecule a charge density    
and a spin. For the general case $\Lambda > 0$, the iron atom in the heme molecule does not have a fixed spin and valence. This is essentially related
to the fact that the iron density and spin are not global observable, and hence they can fluctuate strongly.

We emphasize that the valence fluctuations are not only captured by DMFT, in particular local observables also fluctuate
in density functional theory. However the scheme above, only valid for a many body description (Fock space), does not apply at
the DFT level and a different definition of the entropy needs is called for. To characterise the entropy 
in DFT, the so-called \emphasize{linear entropy} definition needs to be used \cite{linear_entropy_pure_state}. The linear
entropy of the reduced density matrix $\rho_{red}$,
\begin{equation}
 L =  Tr \left( \hat \rho_{red} \right) -   Tr \left(  \hat \rho_{red}^2  \right) 
\end{equation}
may be used as a measure of entanglement for a pure state. Since the reduced density matrix is normalized ($ Tr \hat \rho_{red}=1$), we 
obtain:
\begin{equation}
 L =  1 -  Tr \left( \hat \rho_{red}^2  \right) 
\end{equation}
With this definition of the entropy, suitable for DFT calculations, we found that $L=0.757$ in FeP-p and $L=0.765$ in Fep-d.
The entropy at the DFT level is hence much smaller than the DFT+DMFT entropy, which is in the range $\Lambda=2.5-4.2$.
This is explained by the absence of the many-body excitations induced by the correlations ($U$ and $J$) which contribute
significantly to the entropy.

We note that the linear entropy is an approximation of the Von Neumann entropy, which is often used in DFT calculations since it does not
require a direct inversion of the full density matrix $\rho$. 
Although in principle the DFT entropy obtained for the exact density matrix $\rho$ would capture the effect of quantum entanglement
at zero temperature, in practice approximation are made for the exchange energy functional E$_{xc}$, which in general
do not fully capture quantum fluctuations (the extent to which DFT reproduces the entanglement can be studied in simple problems
where the exchange functional is known exactly, such as in the Hooke's atom \cite{linear_entropy_pure_state}). 

A simple many body version can be constructed from the DFT reduced density matrix $\rho_{red}$
in the spirit of the configuration interaction approach, where we have
many body states.
In the reduced subspace, if we neglect strong off-diagonal components in the DFT reduced density matrix $\rho_{red}$, 
the statistical operator is diagonal, and we can associate to each orbital a many body configuration with a
statistical weight based on the occupation density of this orbital $n$ and on the averaged double occupancy of this orbital $d$.
In particular, we associate to each orbital $\alpha$ \cite{entropy_many_body_state_hand}:
\begin{equation}
\hat \rho_{\alpha} = p_0 |0\rangle \langle 0| + \sum\limits_{\sigma}{ p_1 | \sigma \rangle \langle \sigma |} + p_2 |2 \rangle \langle 2 |
\end{equation}
where $|0\rangle$,$ |\sigma\rangle$,$ |2\rangle$ denotes respectively no electron, an electron with spin $\sigma$, and two electrons in the orbital $\alpha$.
A natural estimation of $p_0$,$p_1$ and $p_2$ is: 
\begin{itemize}
\item
$p_0 = 1- n +d$
\item
$p_1 = n/2 - d$
\item
$p_2 = d $
\end{itemize} 
Because of its diagonal form the local von
Neumann entropy is given by $\Lambda = - \sum\limits_{a}{p_a \text{ln} p_a}$, with $a=0,\uparrow,\downarrow,2$
and the total entropy for the five orbitals $\alpha$ can be obtained similarly. 
To obtain an upper bound on the entropy, we can consider the case without correlations $U=0$ and 
assume $d=n^2/4$. We find $\Lambda=4.72$ for FeP-p and $\Lambda=4.76$ for FeP-d. These values
are of the same order at the maximum of the entropy obtained with DMFT in the regime $J \approx 0.8$eV.

In particular, in the simple argument
above, we assumed that the probabilities associated with every orbital $\alpha$ were independent (no correlations). This simple
picture breaks in the presence of a finite $U$. Indeed, we find that for $J=0$eV and $U=4$eV the entropy is much smaller. 
A large repulsion quite generally reduces the entropy and orders the charge \cite{entropy_many_body_state_hand}.

In conclusion, to take into account many body effects in the presence of correlations in the iron atom ($U$ and $J$), a consistent
method is required. DMFT provides a framework to carry out these calculations without any add-hoc assumptions.

 \section{Methodology: spectral functions resolved in real-space and energy, the optical conductivity and the total energy}
 
The Lanczos DMFT solver used in our study \cite{our_paper_cpt_ed} readily provides 
information on the real frequency axis.
For example, the NGWF-resolved spectral density 
$ \rho_{\alpha \beta} \left(\omega \right)$  is defined, via the
 retarded Green's function obtained at  
 DMFT self-consistency, $\bold{G}$, by
 \begin{equation}
  \label{density}
 \rho^{\alpha \beta} \left(\omega \right)
 =\frac{  G^{  \alpha \beta} \left(\omega \right)
 - G^{ \dagger \alpha \beta} \left(\omega\right) 
  }{ 2i\pi }.
 \end{equation}
 We note that the Green's function has poles on the real axis, such that
 it is in principle not possible to define $G(\omega)$. Hence, the Green's function is defined
 with a small but finite broadening factor $i\eta$, and $G(\omega+i\eta)$ is used in the formula above.
 The small shift $\eta$ plays the role of a broadening factors, similarly to the broadening used in DFT
 to compute the density of states at zero temperature.
 
The DFT+DMFT density of states is then given 
by the imaginary part of the 
many-body density matrix, so that 
 \begin{equation}
 \rho \left( \omega \right) =  \Im \left[ 
  \rho^{\alpha \beta} \left(\omega \right) \right]  S_{\beta \alpha} .
 \end{equation}
 The real-space representation of the \emphasize{Fermi density}, the spectral density at the
 Fermi level is given by 
\begin{align}
 n_F \left( \mathbf{r}\right) = \phi_\alpha \left( \mathbf{r}\right)  
 \Im \left[   \rho^{\alpha \beta} \left(\omega \rightarrow \mu \right)  \right] 
 \phi_\beta \left( \mathbf{r}\right) .
 \end{align}

The theoretical optical absorption was obtained in DFT+DMFT 
within the linear-response regime (Kubo formalism),
including the effect of non-local pseudopotentials 
on the velocity operator matrix elements~\cite{optics_onetep_non_local_potential_commutators},
in the \emphasize{no-vertex-corrections} approximation~\cite{millis_optical_conductivity_review}, 
where it is given by:
\begin{align}
 \sigma(\omega) ={}&  \frac{2 \pi e^2 \hbar }{\Omega }  \int  
 d \omega' \;
 \frac{  f( \omega'-\omega)-f \left( \omega' \right)}{\omega} \\
 {}&
 \times 
 \left( \bold{\rho^{\alpha \beta}} \left( \omega' - \omega \right)  \bold{v}_{\beta \gamma}  \bold{\rho^{\gamma \delta}} \left(\omega' \right) \bold{v}_{\delta \alpha} \right), \nonumber
\end{align}
and the factor of two accounts for spin-degeneracy, 
$\Omega$ is the simulation-cell volume, $e$ is the electron charge,
$ \hbar $ is the reduced Planck constant, 
$f \left( \omega \right)$ is the Fermi-Dirac distribution, 
and the many-body
density matrix $\bold{\rho}^{\alpha \beta}$ is that
defined by equation (\ref{density}).
We note that the optical conductivity will be smoothed out by the broadening parameter $i\eta$ used to obtain
the Green's function on the real axis. In practice, $i \eta$ is chosen to be small enough such that the important features
in the optics can be identified. 

The matrix elements of the velocity operator along one of the direction $\bold{e_x},\bold{e_y},\bold{e_z}$, 
denoted as $\bold{v}^j_{\alpha \beta}$ ($j=(x,y,z)$) are given by
\begin{equation}
\bold{v}^j_{\alpha \beta}= 
- \frac{i \hbar }{ m_e }
\langle \phi_\alpha \rvert  \bold{\nabla}_j \lvert \phi_\beta \rangle
+ \frac{i}{\hbar} 
\langle \phi_\alpha \rvert  
\left[ \hat{V}_{nl}, \mathbf{r} \right] \lvert \phi_\beta \rangle.
\end{equation} 
This expression is general to the nonorthogonal 
NGWF representation~\cite{optical_conductivity_non_orthogonal_basis},
used in this work,
where the contribution to the non-interacting  
Hamiltonian due to the non-local part of the 
norm-conserving pseudopotentials,~\cite{PhysRevB.84.165131,
PhysRevB.44.13071}  
represented by $\hat{V}_{nl}$, is included. It is worth noting that we 
do not invoke the Peierls substitution~\cite{millis_optical_conductivity_review}.

Since optical experiments on heme are typically carried out in liquid or 
gas phases, therefore  
the isotropic part of the optical conductivity tensor:
\begin{equation}
\sigma \left( \omega \right) =
\sum_i \sigma_{ii} \left( \omega \right)/3,
\end{equation}
is of primary interest. Finally, we note that 
optical transitions between hybridised d orbitals (d-d transition type) are forbidden by optical selection rules if the hybridization of the d and p-orbitals happens in a perfect octahedral symmetry , such as in FeP-p.

Once the self energy matrix is obtained, it can be used to correct the DFT total energy with
the DMFT correction. The total energy is estimated by the functional \cite{dmft_dft_total_energy,dft_dmft_total_energy_charge_self_consistence}:
\begin{equation}
E = E_{LDA}(\rho(\bold r)) - \sum\limits_{\bold k \nu}{' \epsilon _{\bold k \nu}} + \text{Tr} \left[  \hat H_{LDA} \hat G \right]  + \langle  \hat H_U  \rangle - E_{DC} 
\end{equation}
where $\hat H_U$ indicates the many body interaction vertex of the DMFT, 
and the primed sum is over the occupied states. 
The symbol "Tr" indicates the one-electron trace for a generic representation and the sum over the Matsubara frequencies $i\omega_n$ for finite-temperature many-body formalism. 

We notice that the total energy within the LDA+DMFT scheme is not simply the expectation value of this Hamiltonian but it consists of several terms, in analogy to the expressions of the usual DFT. 
The first term $E_{LDA}$ contains four different contributions, namely, the ones due to the external potential, the Hartree potential, the exchange-correlation potential, and the sum of the Kohn-Sham eigenvalues. 
However in the spectral density-functional theory, the Kohn-Sham eigenvalues should be reoccupied with respect to the description given by the total GreenÕs function. 
Then we should remove the bare Kohn-Sham eigenvalues sum
(second term) and substitute it with $\text{Tr} \left[ \hat H_{LDA} G \right]$ (third term).

The interaction term $\hat H_U$ can be obtained with the the Galitskii-Migdal formula \cite{total_energy_dft_dmft}:
\begin{equation}
\langle \hat H_U \rangle = \frac{1}{2} \text{Tr} \left[  \hat\Sigma \hat G   \right]  
\end{equation} 
Where $\hat \Sigma$ ($\hat G$) is the self-energy (Green's function) matrix in the NGWFs representation.  
We note that both the self energy and the Green's function are slowly decaying function, hence the trace over matsubara frequencies has to be done
with care. In this work, we followed the steps of Refs. \onlinecite {dmft_dft_total_energy,dft_dmft_total_energy_charge_self_consistence}.

Finally, the double-counting correction $E_{DC}$ must be introduced, since the contribution of interactions between the correlated orbitals to the total energy is already partially included in the exchange-correlation potential derived from the DFT.  The most commonly used form of the double-counting term is \cite{hund_coupling_averaged_double_counting}:
\begin{equation}
E_{DC} = \frac{U_\mathrm{av}}{2} N_d \left( N_d - 1 \right)  - \frac{J}{2} \sum\limits_\sigma{N_{d\sigma} \left( N_{d\sigma}-1 \right)}
\end{equation}
where $U_\mathrm{av}$ is the averaged Coulomb repulsion defined in equation (\ref{uav}).

  
\section{Validation of the methodology}

In order to validate our approach, we have studied the dependence of our results with the other parameters of the theory, e.g. the temperature $T$
and the Coulomb repulsion $U$. 

\begin{figure*}
\begin{center}
\includegraphics[width=2 \columnwidth]{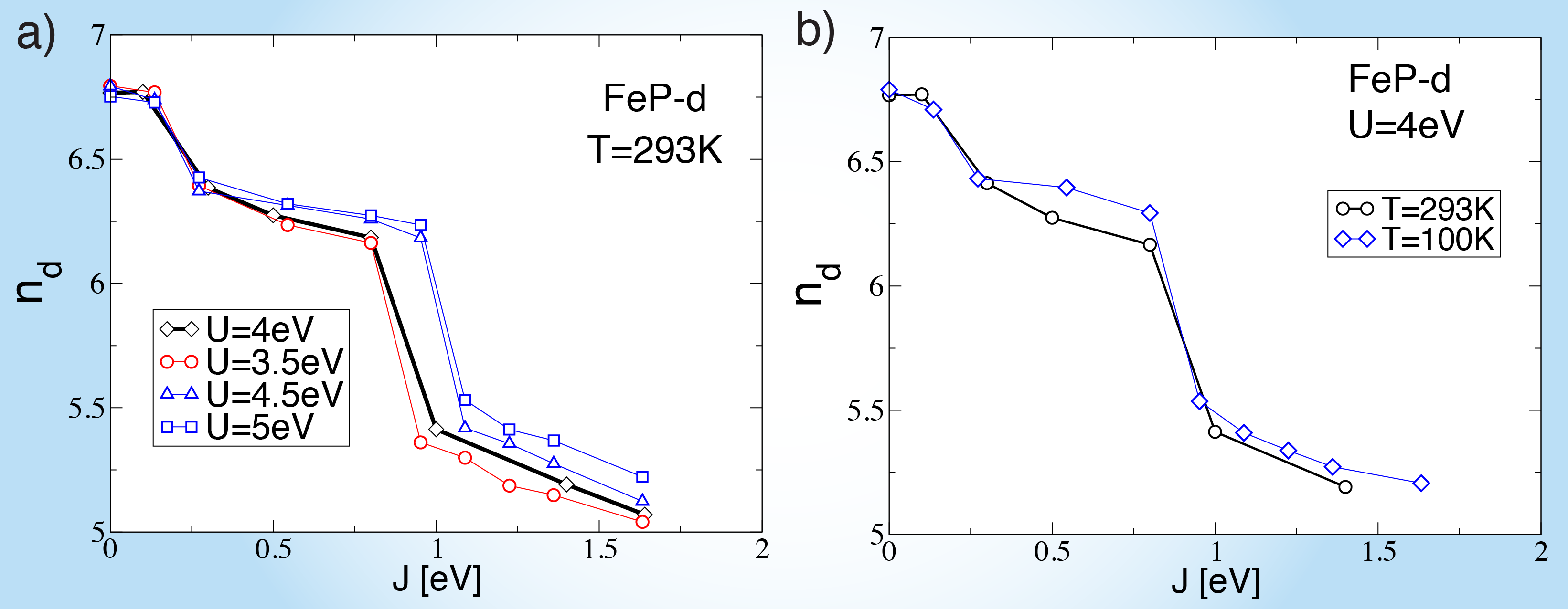}
\caption{ (Colors online) a) Iron occupation $n_d$ as a function of the Hund's coupling $J$ for different Coulomb repulsion $U$ for FeP-d. The data
discussed in the paper are reproduced here for comparison ($U=4$eV, bold line). The sharp transition in the iron density is present for all repulsions. b) Iron occupation $n_d$
for two different temperatures $T=100$K (diamonds) and $T=293$K (circles).}
\label{fig7}
\end{center}
\end{figure*}

We show in Fig.~\ref{fig7}.\textbf{a} the iron occupation $n_d$ as a function of $J$ for different values of $U=3.5$eV, $U=4$eV, $U=4.5$eV, and $U=5$eV for FeP-d. 
We find that the results are qualitatively similar, for a $1.5$eV variation of U ($3.5-5$eV), although the critical value $J$ associated with the sharp drop
of the iron density is slightly shifted from $J=0.8$eV to $J=1$eV. We note however that for $J=0$eV, the density is only very marginally affected as a function
of $U$. The latter suggest that strong modifications of the physical properties are induced principally by the Hund's coupling. In our work, we chose $U=4$eV, in agreement
with earlier DFT studies \cite{heme_marzari}, and we show that although variations in U might give some quantitative modifications at J=0, there is a dramatic change
in the occupation density at finite J, neglected so far in the DFT studies of heme. We also note that it is physically better to take into account the Hund's coupling when dealing
with metallic atoms such as iron, known to have a large Hund's coupling. 

We also studied how our results depend on the temperature T (see Fig.~\ref{fig7}.\textbf{b}). We compared the data obtained at $T=100$K and $T=293$K.
Despite some minor quantitative differences, the sharp transition in the iron density is obtained for the same critical Hund's coupling $J$, showing
that our predictions are valid for a broad range of physically relevant temperatures. This confirms also that quantum fluctuations are mostly dominating 
in this regime of temperature, although we cannot exclude that for $T<100$K we might observe some larger changes. 

In conclusions, we believe that our predictions are not dependent on the details of our model, and we show that the parameter which has a dramatic
impact on the physics in the Hund's coupling. 

\begin{figure*}
\begin{center}
\includegraphics[width=2\columnwidth]{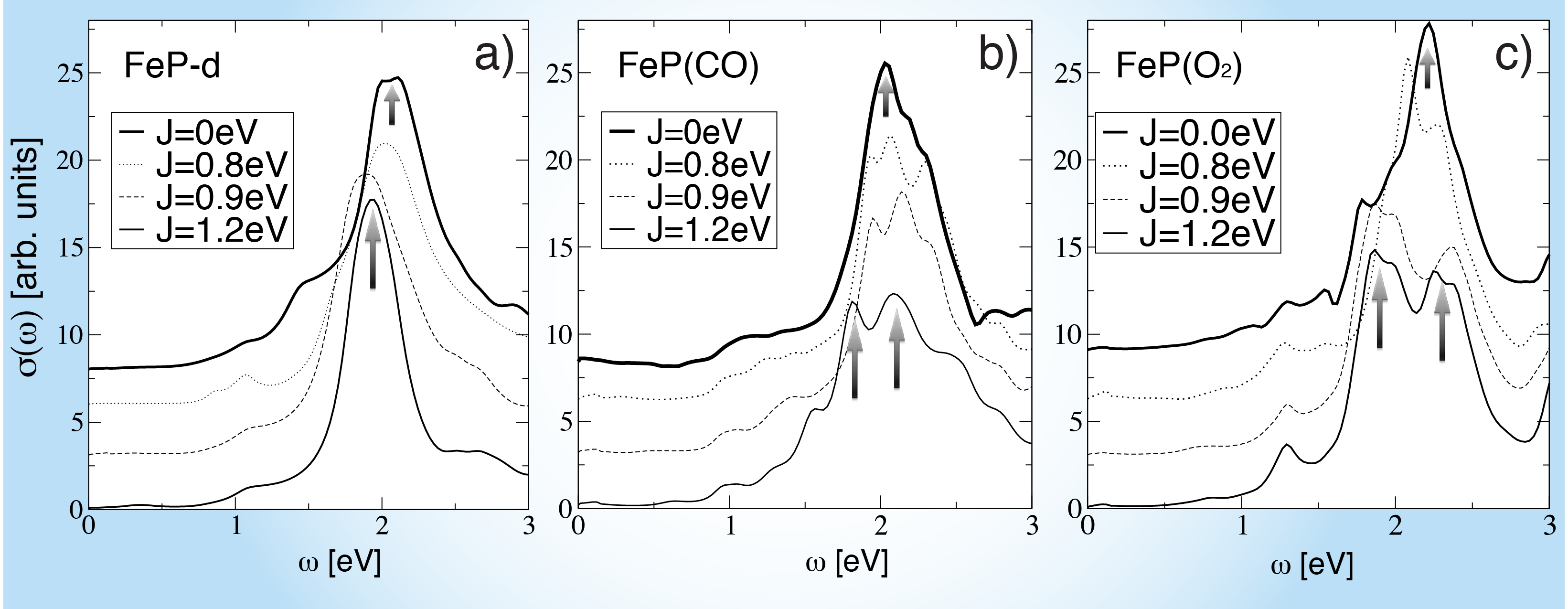}
\caption{ Optical spectra for a) FeP-d, b) FeP(CO) and c) FeP(O$_2$) for different values of $J$ (from top to bottom): $J=0$eV (bold line), $J=0.8$eV (dotted line), $J=0.9$eV (dashed line), $J=1.2$eV (full line).
The arrows highlight the presence of a single peak structure for $J=0$eV, which splits in a double peaked structure as $J$ increases. For clarity, the curves were shifted on the vertical axis.}
\label{fig6}
\end{center}
\end{figure*}

To support further the justification that the Hund's coupling is important for heme, 
we now turn to the discussion of the optics for FeP-d (Fig.~\ref{fig6}.\textbf{a}), FeP(CO) (Fig.~\ref{fig6}.\textbf{b}) and FeP(O$_2$) (Fig.~\ref{fig6}.\textbf{c}) as a function of $J$.
For unligated heme (Fig.~\ref{fig6}.\textbf{a}), we do not find any qualitative difference in the optical spectra as $J$ is increased from $0$eV to $1.2$eV. However,
for both FeP(CO) and FeP(O$_2$), we find that increasing $J$ gives a double peaked structure at $\omega \approx 2$eV, which is also observed in experiments (see inset
of the Fig. 3 of the paper). This double peaked structure, absent for $J=0$eV, suggests also that $J$ is important to describe the heme molecule.
We note that the best fit to the experiments is obtained for $J=0.9$eV, and hence we can furthermore support that the Hund's coupling is consequent in heme based on phenomenological grounds. 

 
 \section{Effect of the Hund's coupling on the orbital polarisation and binding to ligands}
 
 As shown in the Fig.~3\textbf{a} of the paper, the effect of the Hund's coupling is to reduce the
 difference in the FeP binding energies to CO and O$_2$ ($\Delta \Delta E$). In this section we
 give more details regarding the mechanism underlying the reduction of $\Delta \Delta E$ induced by $J$.

\begin{figure}
\begin{center}
\includegraphics[width=0.6\columnwidth]{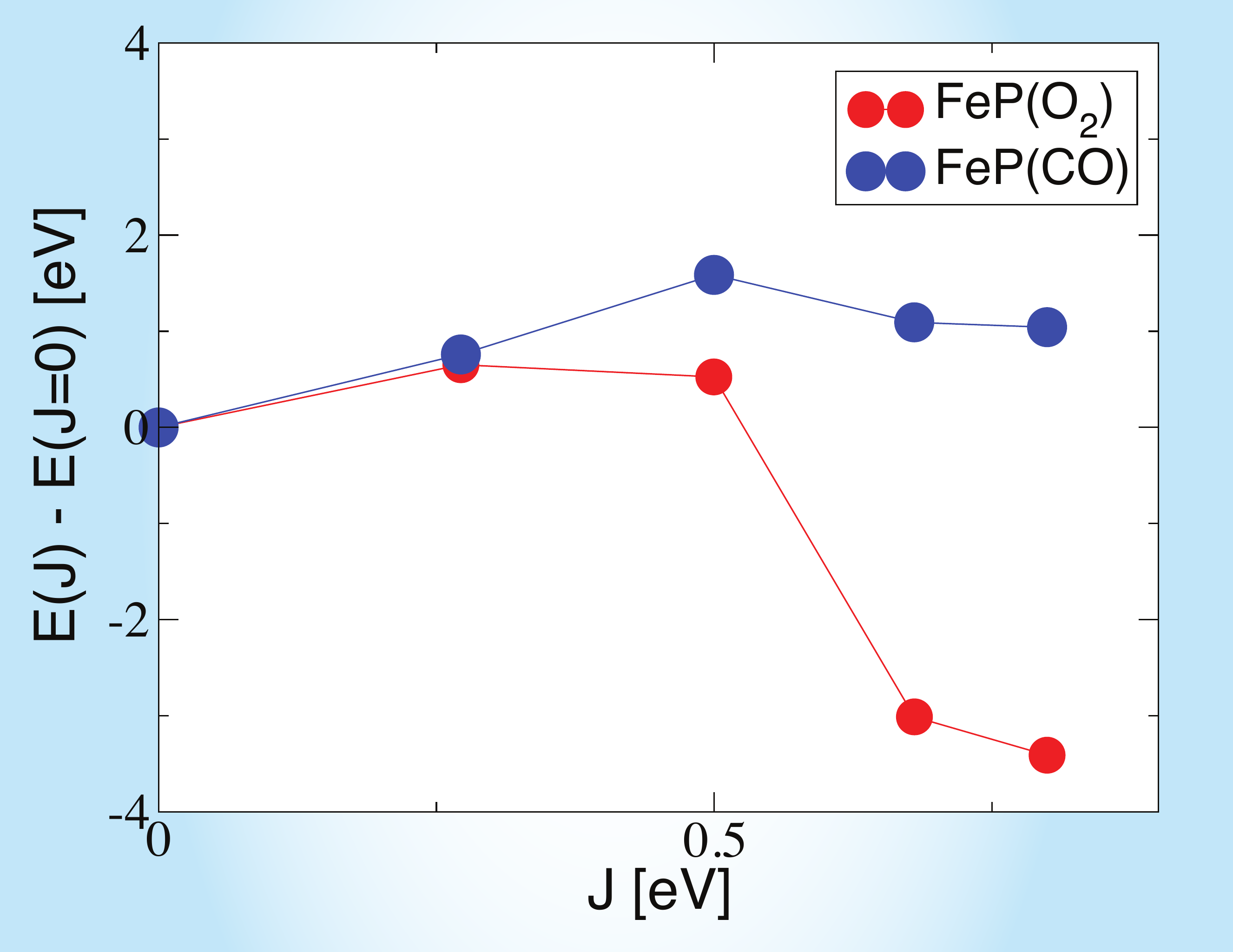}
\caption{ 
{\bf Energy: }
variation of the total energy $\Delta E=E(J)-E(J=0)$ of FeP(O$_2$) (red circles) and FeP(CO) (blue circles)
as a function of the Hund's coupling. While there is a drop in the energy for FeP(O$_2$) at $J \approx 0.8$, the energy
of FeP(CO) is only weakly affected by the Hund's coupling.}
\label{fig8}
\end{center}
\end{figure}

 In Fig.~\ref{fig8} we show the energy variation for FeP(CO) and FeP(O$_2$) as the Hund's coupling $J$
 increases. We find that while the energy of FeP(CO) is only weakly affected by $J$, there is a sharp
 drop of the energy of FeP(O$_2$) at $J \approx 0.8$ eV. 
 
 \begin{table}[ht]
\begin{tabular}{|c|c|c|c|c|c|c|}
\hline
           &  $J$   &   d$_{x^2-y^2}$  &  d$_{3z^2-r^2}$&    d$_{xz}$&    d$_{xy}$&    d$_{yz}$  \\
\hline
\hline
FeP        &  0   &   0.85 & 1.86&   1.24&   1.98&   0.82 \\
FeP        &  0.8 &   1.10 & 1.75&   1.08&   1.14&   1.08 \\
\hline
FeP(CO)    &  0   &   1.06 & 0.86&   1.99&   1.06&   1.99 \\
FeP(CO)    &  0.8 &   1.14 & 1.33&   1.16&   1.05&   1.85 \\
\hline
FeP(O$_2$) &  0   &   0.72 & 1.82&   1.25&   1.87&   1.28  \\
FeP(O$_2$) &  0.8 &   1.03 & 1.07&   1.18&   1.97&   1.09  \\
\hline
\hline
\end{tabular}
\caption{Average occupations $n_d^\alpha$ of the iron d orbitals for FeP, FeP(CO) and FeP(O$_2$), for J=0 and J=0.8.}
\label{table2}
\end{table}

\begin{table}[ht]
\begin{tabular}{|c|c|c|c|c|c|c|}
\hline
          &   $J$   &   d$_{x^2-y^2}$ &    d$_{3z^2-r^2}$  &   d$_{xz}$  &   d$_{xy}$  &   d$_{yz}$  \\
\hline
\hline
FeP       &   0   &   0.30&    0.15 &   0.28 &   0.05 &   0.28 \\
FeP       &   0.8 &   0.31&    0.20 &   0.39 &   0.38*&   0.39 \\
\hline
FeP(CO)   &   0   &   0.37&    0.37 &   0.03 &   0.48 &   0.03 \\
FeP(CO)   &   0.8 &   0.37&    0.40 &   0.46*&   0.48 &   0.19 \\
\hline
FeP(O$_2$)&   0   &   0.35&    0.20 &   0.34 &   0.17 &   0.34 \\
FeP(O$_2$)&   0.8 &   0.37&    0.43*&   0.45 &   0.08 &   0.46 \\
\hline
\hline
\end{tabular}
\caption{Internal magnetic moment of the iron d orbitals for FeP, FeP(CO) and FeP(O$_2$), for J=0 and J=0.8. The internal
orbital magnetic moment is obtained by
$S_\alpha=\sqrt{\langle \bold{S_\alpha}\bold{S_\alpha}\rangle}$, where $\alpha$ is an index
for the d orbital and $\bold{S_\alpha}$ is the spin operator of the orbital $\alpha$. The star highlights the orbital with
the largest moment increase.}
\label{table3}
\end{table}

\begin{figure}
\begin{center}
\includegraphics[width=0.8\columnwidth]{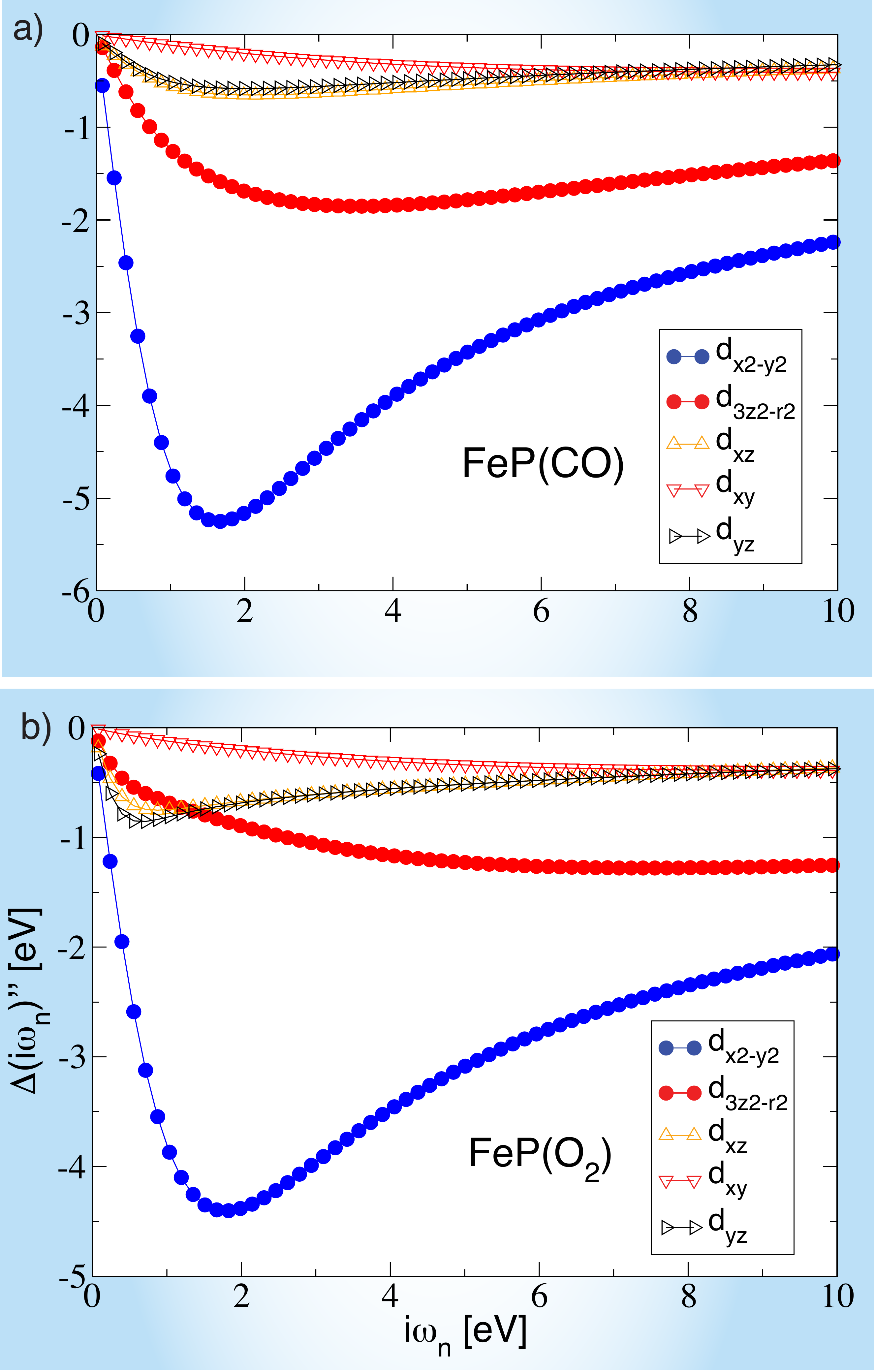}
\caption{ 
{\bf Iron dynamical hybridization:}
Imaginary part of the hybridization $\Delta''$ as a function of the matsubara frequencies $i\omega_n$ of the iron d orbitals for a) 
FeP(CO) and b) FeP(O$_2$). The $e_g$ orbitals have the largest hybridization.
Note that the hybridization is independent of the many body parameters $U$ and $J$.}
\label{fig9}
\end{center}
\end{figure}

 The latter can be understood in terms of orbital polarisations. In particular, we compare  
 in Table~\ref{table2} the electron occupation of the iron d orbitals for $J=0$ and $J=0.8$ eV.
 We note that the orbital occupations are not integer numbers. Indeed, in the presence
 of hybridisation (see Fig.~\ref{fig9}.\textbf{a,b}), the atomic iron densities (local observables) are not conserved quantum numbers, 
 and can take fractional values, which is a signature of valence fluctuations. 
 
 However, due to the form of the hybridization of the iron atom in FeP, the 
 orbitals (\dz,\dxy) are almost filled and hence do not fluctuate (Table~\ref{table2}) in the absence of the Hund's coupling.
  We also find that for FeP(CO) and FeP(O$_2$), the orbital
 which are almost full are respectively the (\dxz,\dyz) and the (\dz,\dxy) orbitals.
 
 The main effect of the Hund's coupling $J$ is to increase the magnetic moment of the iron atom, which in
 turn yields a concomitant reduction of the iron density $n_d$. For FeP, there is a reduction of the iron density 
 of 0.52~e (see Table~\ref{table1}), induced by the build up of the self energy at the iron site due to the Hund's coupling and the Coulomb repulsion,
 and a reduction of 0.25~e of the neighbour N atoms. The charge is transferred to the two hydroxyl chains which are electronegative
 (each hydroxyl chain contains two O atoms).
 
\begin{table}[ht]
\begin{tabular}{|c|c|}
  \hline
 atom   & $\Delta n(r) $   \\  
  \hline
Iron d orbitals     &  -0.52 \\
Nitrogen ring     &  -0.25 \\
hydroxyl groups     &  +0.77 \\
   \hline
\end{tabular}
\caption{Variation of the charge $\Delta n(r)=n(r,J=0.8)-n(r,J=0)$ in FeP induced by the Hund's coupling.}
\label{table1}
\end{table}

We find that in FeP(CO), the orbital
mostly affected by an increase of $J$ is the \dxz: the occupation of the latter is reduced from 2~e ($J=0$) down
to 1.16~e ($J=0.8$eV). For FeP(O$_2$), the orbital which is strongly affected by $J$ is the \dz orbital: 
the occupation of the latter
is reduced from 1.8~e down to 1.07~e. Moreover, we find
that the reduction of the occupation of \dxz in FeP(CO) and the reduction of the occupation of \dz
in FeP(O$_2$) are consistent with a build up of the magnetic moment in the latter orbitals  (see Table~\ref{table3}).
We note that both the Coulomb repulsion $U$ and the Hund's coupling $J$ have an importance here, and promote
different many body configurations.
The Coulomb repulsion $U$ tend to suppress doubly occupied many body configurations, 
whereas the Hund's coupling tends to generate many body configurations with aligned spins. 
In particular, a physical constraint on the Hund's coupling $J$ is $J<U$ (see Ref.~\onlinecite{imada_mott_review}). 

The reduction of the charge of the \dxz orbital in FeP(CO) and of the \dz orbital in FeP(O$_2$) is
expected to reduce the Coulomb energy, which penalises doubly occupied orbitals. 
Moreover, the charge reduction is also expected to yield an optimisation of the kinetic energy,
since transfer of electrons to a filled orbital are impossible.
The latter is however not effective in FeP(CO), since the \dxz orbital does not hybridise strongly (see Fig.~\ref{fig9}.\textbf{a}) and hence
no drastic change is expected in the kinetic energy,
but the latter is expected to be important for FeP(O$_2$) since the \dz orbital hybridise significantly (see Fig.~\ref{fig9}.\textbf{b}). This accounts for
the further energy reduction in FeP(O$_2$) at $J \approx 0.8$eV, not present in FeP(CO), as observed in Fig.~\ref{fig8}.

\section{Quantum dynamics: transient magnetic response simulations}

\begin{figure}
\begin{center}
\includegraphics[width=\columnwidth]{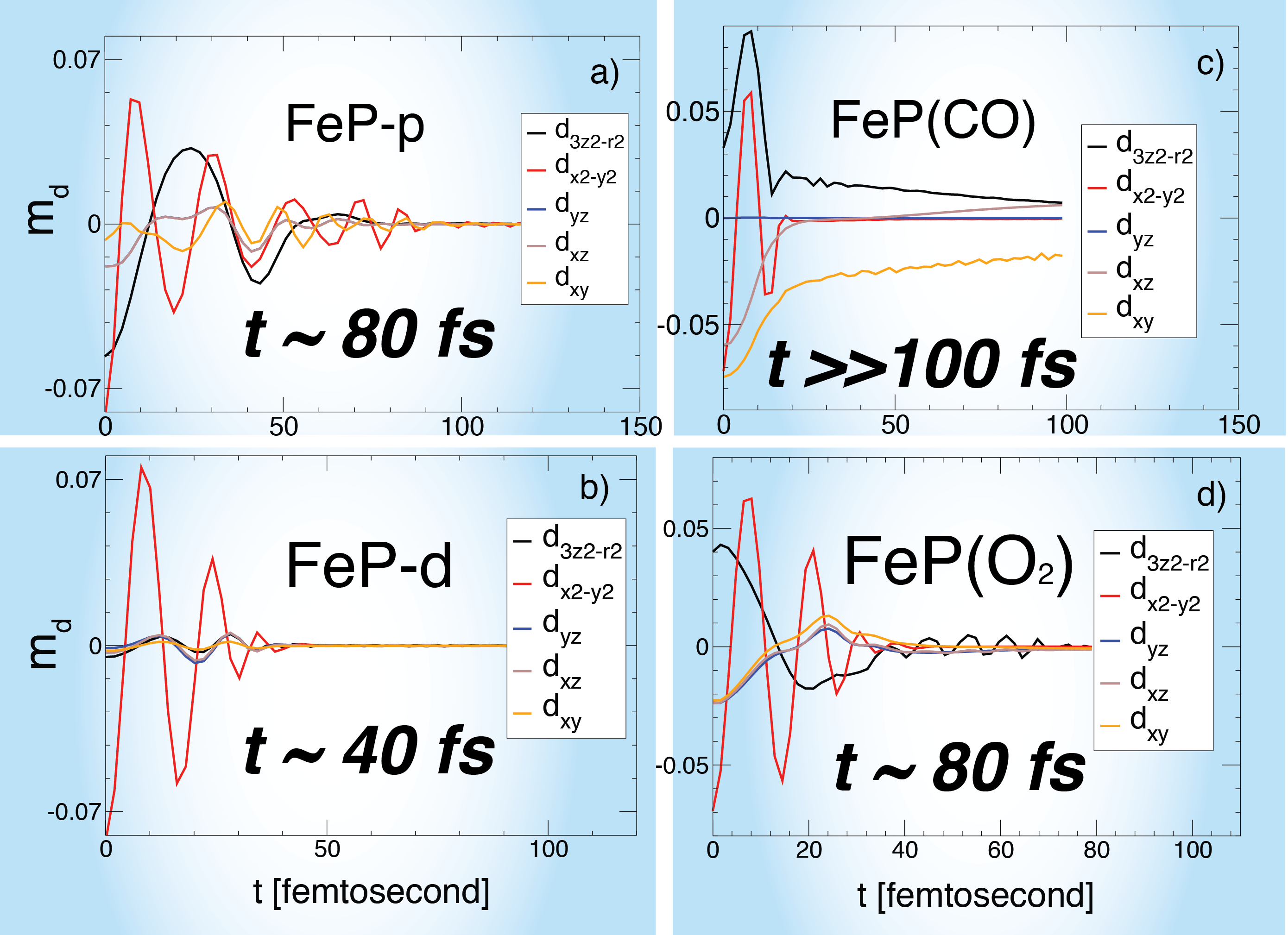}
\caption{ 
{\bf Transient magnetic response: }
Time dependence of the subspace magnetization 
 $m_d = ( n_d^\uparrow - n_d^\downarrow ) \hbar / 2$, after an
initial quench in a magnetic dipole field, for heme species 
a) FeP-p, b) FeP-d, c) FeP(CO) and d) FeP(O$_2$).}
\label{fig4}
\end{center}
\end{figure}
We analyzed  the quantum dynamics 
of heme by means of out-of-equilibrium calculations 
performed using a recently-developed implementation
of the Keldysh formalism adapted for Lanczos DMFT
solvers~\cite{keldysh_ed_lanczos}. 
Our calculations allow us to directly tackle 
the issue of dynamical relaxation after a quantum quench; 
the results are shown in Fig.~\ref{fig4}.
Specifically, we applied a quench 
in molecular spin polarization at $t=0$, by perturbing 
the ground state with a magnetic dipole impulse 
applied to the 
iron $3d$ impurity subspace and 
thereafter studying the transient behavior of the 
spin polarization.
The heme molecule plays the role of a dissipation bath, such that
after a long enough time the system relaxes to the equilibrium.
The relaxation times for unligated deoxyheme, at  
$J=0.8$~eV, were found to be
approximately $t=80$~fs (Fig.~\ref{fig4}.\textbf{a})
 and $t=40$~fs (Fig.~\ref{fig4}.\textbf{b}) 
 for FeP-p (unligated, planar) and FeP-d (unligated, domed),
  respectively. 
We find that in the FeP-d system, 
only the \dx orbital is strongly spin polarized by the quench, 
whereas, in FeP-p, the \dz orbital
is also partially occupied in the ground-state and
hence also participates in the magnetic response, 
yielding  transient oscillations that are longer-lived overall. 
Our result suggests that the dome-shaped configuration 
protects heme from external perturbations to some extent,
reducing the hybridization between 
the \dx and \dz orbitals by further lifting their degeneracy. 
For FeP(CO) (Fig.~\ref{fig4}.\textbf{c}), we observe a 
very long relaxation time (beyond our numerical reach);
magnetically perturbed FeP(CO) becomes trapped 
in a metastable state with a slowly-decaying internal magnetization.
Finally, for FeP(O$_2$), we find an intermediate 
relaxation time of $t=80$~fs (Fig.~\ref{fig4}.\textbf{d}),
and transient response rather different to that of FeP(CO),
a difference which appears to be primarily due to a 
magnetic equivalence among the \tg orbitals of FeP(O$_2$) which
is absent in FeP(CO).
These results highlight the surprisingly long time-scale 
and system dependence associated with quantum spin relaxation.
Photo-excitation of carboxyheme, FeP(CO),
has previously been  discussed in the context of
both femtosecond spectroscopy~\cite{heme_photolysis} and
time-resolved resonance Raman 
spectra~\cite{heme_raman_dynamics_femto},
where the presence of excited states with a lifetime
of 300~fs was attributed to iron-to-porphyrin 
charge-transfer excitations.
We suggest, based on the results of our out-of-equilibrium 
simulations (Fig.~\ref{fig4}),
that  circularly-polarized two-photon spectroscopy
 may be used as a sensitive probe for
  the geometric configuration and ligation state of heme.


%